\documentclass[preprint,12pt]{revtex4-1}
\pdfoutput=1
\usepackage{bm,bbm,graphicx,epstopdf,setspace}
\usepackage{anysize}
\usepackage{fancyhdr}
\usepackage{graphicx,subfigure,wrapfig,color,enumerate}
\usepackage{amssymb,amsfonts,amsmath,psfrag,latexsym}
\usepackage{mhchem,rsphrase,cancel,soul}

\newcommand\reaction[1]{\begin{equation}\ce{#1}\end{equation}}
\newcommand\reactionnonumber[1]%

\def\be{\begin{equation}}
\def\ee{\end{equation}}
\def\bea{\begin{eqnarray}}
\def\eea{\end{eqnarray}}

\def\r{{\bf r}}

\def \kbt{$k_{\rm B}T$}

\def \nmcs{$N$/MCS}
\begin{document}

\title{Active remodeling of chromatin and implications for  in-vivo folding}

\author{N. Ramakrishnan}
\affiliation{Department of Bioengineering,  University of Pennsylvania, Philadelphia, PA-19104, USA}
\affiliation{Department of Physics, Indian Institute of Technology Madras, Chennai 600036, India}
\author{Kripa Gowrishankar}
\affiliation{Raman Research Institute, C.V. Raman Avenue, Bangalore 560080, India}
\affiliation{Azim Premji University, Hosur Road, Bangalore 560010, India}
\author{Lakshmi Kuttippurathu}
\affiliation{Department of Physics, Indian Institute of Technology Madras, Chennai 600036, India},
\author{P. B. Sunil Kumar}
\affiliation{Department of Physics, Indian Institute of Technology Madras, Chennai 600036, India}
\author{Madan Rao}
\affiliation{Raman Research Institute, C.V. Raman Avenue, Bangalore 560080, India}
\affiliation{Simons Centre for the Study of Living machines, National Centre for Biological Sciences (TIFR), Bellary Road, Bangalore 560065, India}

%\contributor{Submitted to Proceedings of the National Academy of Sciences of the United States of America}

\begin{abstract}
Recent high resolution experiments have provided a quantitative description of the statistical properties of  interphase chromatin  at large scales.  These findings have stimulated a search for generic physical interactions that give rise to such specific statistical conformations. Here, we show that an  active chromatin model of {\it in-vivo} folding, based on the interplay between polymer elasticity, confinement, topological constraints and active stresses arising from the (un)binding of ATP-dependent chromatin-remodeling proteins gives rise to steady state conformations consistent with these experiments. Our results lead us to conjecture that the chromatin conformation resulting from this active folding optimizes information storage by co-locating gene loci which share transcription resources.
\end{abstract}

%\keywords{chromatin | fractal polymer |  active stresses | contact probability | Peano or Hilbert  curves}

%\abbreviations{CRP, Chromatin-remodeling protein; DMC, Dynamical Monte Carlo}

%\section{Significance Statement}
%Recent experimental studies on the conformations of  interphase chromatin are providing fresh insight into the physical principles of {\it in-vivo} folding of chromatin at large scales in a variety of model  systems. We show that a generic  interplay between polymer elasticity, confinement, topological constraints and active stresses arising from ATP-dependent mechano-chemical remodeling of chromatin can give rise to space-filling conformations at steady state, with properties consistent with these observations. This places active, energy-consuming processes and topological constraints as key determinants of {\it in-vivo} chromatin folding at large scales. We conjecture that generic conformations of chromatin achieved as  a result of this active folding, optimize information storage by co-locating gene loci which share transcription resources.

\maketitle

\section{Introduction}
While there has been substantial progress in understanding the  molecular basis of structural organization of chromatin at smaller scales, a statistical description of the 3-dimensional (3d) organization at large scales has until recently proved a challenge. Studies of interphase chromatin architecture using FISH and Hi-C techniques 
on a scale of a megabase in three different systems, human~\cite{LiebermanAiden:2009jz}, mouse~\cite{Zhang:2012iu} and drosophila chromosome 3R~\cite{Sexton:2012cn}, have now provided striking and  consistent results on the statistics of packing and looping, such as   (i) the ``size'' of the chromosome measured by its radius of gyration $R_G^{2} \propto N^{2\nu}$, where $\nu \approx 1/3$ ($N$ is the number of base pairs), i.e., over the largest scale    chromatin is compact,   (ii) the ``size'' of the chromatin over a contour length $s > s_{min}$, also scales as $R_G^{2} \propto s^{2\nu}$, with the same $\nu$, suggesting that at scales beyond $s_{min}$ the chromatin is self-similar and   compact, and   (iii) the contact probability $P(s)$  is independent of the nucleotide sequence but  dependent of its state of activity (thus euchromatin and heterochromatin display different loop statistics), with $P(s) \sim s^{-\theta}$,  where $\theta \approx 1$. These statistical properties are clearly inconsistent with an {\it equilibrium globule}~\cite{LiebermanAiden:2009jz}; the proposal was that the statistics of these conformations could be described by a self-similar,  space filling,  {\it fractal} or {\it crumpled} polymer~\cite{Grosberg:1993fj,Mirny:2011cl}. These statistical features do not appear to be entirely universal - in human embryonic stem (ES) cells,  the chromatin is more swollen and the measured contact probability is closer to $P(s) \sim s^{-3/2}$~\cite{wolynes:2015}.

Taken together, these studies reinforce the claim that  chromatin folding at large scales is a problem in polymer physics, as was recognized by~\cite{vandenEngh:1992bi}. This has naturally provoked the question - what are the physical forces or processes that give rise to such a statistical description of the `chromatin polymer' at large scales?   A recent review~\cite{Halverson:2014hg} summarizes the existing polymeric models proposed to explain this 3d chromatin architecture. Models based on purely equilibrium forces arising from polymer elasticity and interactions with DNA-binding proteins~\cite{Nicodemi:2009hb,Barbieri:2012iw,Brackley:2013dt}, show the  spontaneous formation of chromosome co-localization and looping and territory formation via the establishment of `bridging' proteins at two or more chromosomal sites. These models may  however be relevant  to chromatin looping at scales much smaller than those probed by Hi-C, as argued in ~\cite{Halverson:2014hg}.
On the other hand, ~\cite{LiebermanAiden:2009jz,Halverson:2014hg,Rosa:2008bp} propose that the crumpled polymer configuration is a result of a fast collapse of the chromatin structure whose  relaxation  to an equilibrium globule configuration is hindered by topological or steric constraints of the long polymer in a crowded nuclear environment. Here, we show that the observed 3d conformations arise naturally by modeling chromatin as an active semiflexible, self-avoiding polymer confined within the nucleus. Nuclear chromatin is subject to  {\it active} stresses from binding/unbinding of a variety of chromatin-remodeling proteins (CRPs)~\cite{Saha:2006do}, leading to the local release of histones.
%-- the energy cost in their unbinding  is $\approx 40$ \kbt{}~\cite{BrowerToland:2002fr}. 
In this paper, we include the mechanochemistry of the CRPs and  study the Rouse dynamics~\cite{doi:1986bk} of the chromatin using a coarse-grained Dynamical Monte Carlo simulation. For the  active polymer in 3d, we recover the statistical features reported in the Hi$-$C experiments for mammalian chromosomes, both in differentiated and ES cells.

Interestingly, Hi-C studies of the yeast chromatin~\cite{Duan:2010jf}, showed a different scaling, {\it viz}., $R_G^{2} \propto N$ and $\theta \approx 3/2$, and have been taken to be consistent with an  equilibrium globule. Rather than interpret the yeast chromatin as following different folding principles, we ask whether this 
statistical behavior appears as a solution of the same folding strategy as the other cell types investigated; this is consistent with the suggestion that there may be many global attractors on any realistic chromatin folding landscape~\cite{wolynes:2015}. In this spirit, we find that the same active chromatin model can be made to work for the yeast chromosome too, if we recognize that  in these model organisms each chromosome has a centromere attachment  and chromosome ends (telomere regions) are preferentially located next to the nuclear envelope~\cite{Duan:2010jf}. Such multiple attachment sites
will put significant constraints on the excursions of the polymer in configuration space, which in the extreme limit, is equivalent to an effective reduction in spatial dimensionality. As we will see, the active polymer in 2d recapitulates the statistical features of yeast chromatin.
 
The present study should be viewed as part of a growing body of work that recognises the relevance of active processes in the rheology~\cite{Hameed:2012ej,Bruinsma:2014jv,Weber:2012gd,Ghosh:2014ky} and organization~\cite{Talwar:2013bh,Ganai:2014kv} of nuclear matter at different scales.  

\section{Model}
\subsection{Chromatin elasticity and confinement}
The Hi-C experiments revealed the statistics of organization of nuclear chromatin on the scale of around $0.5-10$\,Mb~\cite{LiebermanAiden:2009jz}, with a resolution of 1Mb.  Over this scale, it is appropriate to treat the enclosed chromatin as a coarse-grained semiflexible polymer, where each ``monomer'' is taken to comprise a group of around $20$ neighboring nucleosomes or $4$\,kbp of dsDNA ~\cite{Rosa:2008bp,Halverson:2014hg}. For a statistical description of chromatin over large scales, the arrangement of nucleosomes within each coarse-grained monomer and the sequence dependence of the polymer should be irrelevant. 
To a first approximation therefore, the coarse-grained chromatin is a semiflexible homopolymer with interactions between monomers, nuclear solvent, and confining surfaces or anchoring to specific substrates. Monomer-monomer interactions include steric and bending interactions. Confinement is imposed by a constraint on the enclosed volume $V$ (in 3d), the Lagrange multiplier fixing this constraint is the pressure, $\Delta p$.
As discussed above, chromatin may also be subject to constraints due to anchoring to substrates. 
%Potential examples  of this might include yeast and bacterial chromatin. 
In such cases, the conformations explored by chromatin would be considerably restricted -- in the extreme limit of very strong anchoring to a 2d substrate it is reasonable to model it as an effective polymer embedded in 2d.
\subsection{Active mechano-chemical remodeling of chromatin}
In addition to being subject to thermodynamic forces from polymer elasticity, the chromatin is driven out-of-equilibrium by the binding and unbinding of ATP-dependent CRPs. Chromatin is tightly bound to histones : single-molecule studies have estimated that the histone-chromatin unbinding energy scales are of order $40$ \kbt{}~\cite{BrowerToland:2002fr}. The binding of the ATP-dependent CRPs generate local active stresses leading to the release of histones from the chromatin. The isotropic part of the active stresses contribute to an osmotic pressure, which extrudes the fluid through the polymeric gel.  In addition, the bound CRPs generate anisotropic active stresses - force dipoles - whose normal and tangential components affect polymer bend and stretch (compression) conformations \footnote{This will also lead to twist deformations of the polymer, which we ignore for simplicity.}. 
 Normal stresses result in local bending of the polymer, while tangential stresses result in a local stretch via the release of a few histones. Conversely, when histones bind to naked DNA, they locally compress the chromatin length. Let us assume that a length $\Delta l$ is added when the action of the CRP releases a group of histones (or subtracted when a group of histones bind to chromatin). The CRP-histone complex are thus {\it strain} (both bend and stretch) {\it generators}. Since strain generators are also {\it strain sensors},  the binding-unbinding kinetics of CRPs should be dependent on the local curvature and stretch.
% (or equivalently local projected density, see {\it S.I}). 
 This mechano-chemical (un)binding process, shown in Fig.1b, may be represented as
\reaction{CRP$_u$ +  Chr  ->[k_{+}(\text{Chr})] CRP$_b$ + Chr ->[+\text{ATP}][-\text{hist}] CRP$_u$ + Chr$^+$} 
\reaction{Chr  ->[k_{-}(\text{Chr})][+\text{hist}] Chr$^-$} 
where Chr represents chromatin whose local state of bend and stretch is taken as reference, Chr$^\pm$ represents chromatin which is locally stretched (compressed) with respect to the reference, by an amount $\Delta l$ due to the loss (gain) of a group of histones, and the subscripts $b$ and $u$ refer to the bound and unbound  CRP. The rate $k_{\pm}$ depend on the local state of the chromatin, namely its bend and stretch. Since changes in the local strain of chromatin is felt at larger scales, both the effective ``forward'' \ce{Chr  ->[k_{+}] Chr$^+$} and ``backward'' \ce{Chr  ->[k_{-}] Chr$^-$} reactions are mechano-chemical and involve energy consumption.
\begin{figure*}[!h]
\begin{center}
\includegraphics[width=15cm,clip]{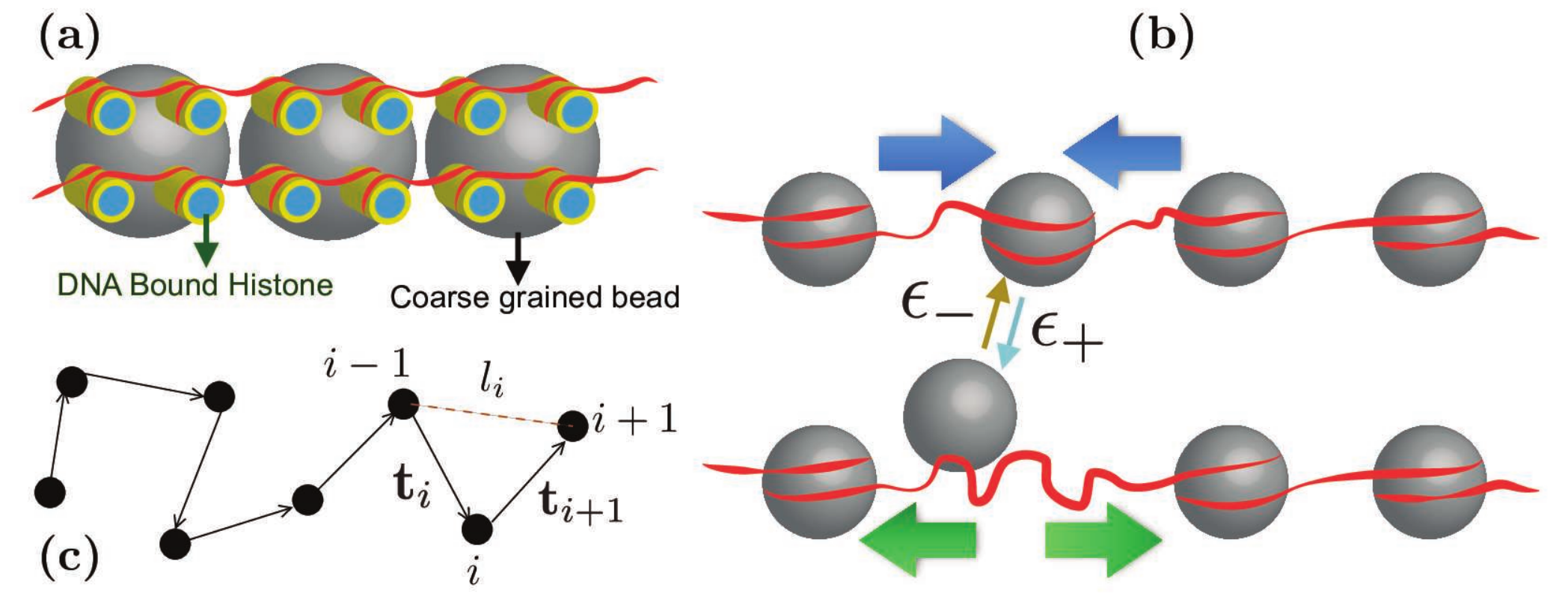}
\caption{{\bf Active mechano-chemical remodeling of chromatin.} (a) Schematic of a section of the nucleosomal complex showing DNA (red strands) wound around a collection of histones (yellow cylinders). The (grey) spheres enclosing the nucleosomes depict the level of coarse-graining used in our study, and corresponds to a group of around $20$ neighboring nucleosomes or $4$\,kbp of dsDNA. ATP-dependent chromatin remodeling proteins (CRPs) bind and unbind on the chromatin and apply active stresses, leading to the local release (or sequestration) of the group of histones within the coarse-grained sphere. (b) At a coarse-grained level, we treat the enclosed chromatin as a semiflexible polymer, where each ``monomer'' is taken to be the grey sphere described above.  Active CRPs apply normal and tangential stresses; the former changes shape and the latter creates local stretch (compression) with respect to the reference polymer (shown by thick green arrows) with rates $\epsilon_{\pm}$. (c) We work in the {\it dual representation}, where the addition and removal of each (black) bead now corresponds to the gain or loss of a free stretch of chromatin length. The tangent vector of the $i^{th}$ segment is denoted by ${\vec t}_i$. This is  the effective semiflexible polymer on which we perform the Dynamical Monte Carlo simulations.}
\label{schematic}
\end{center}
\end{figure*}

\subsection{Dynamical Monte Carlo Simulation (DMC)}
We study the dynamics of the active chromatin using a simple DMC simulation, where the  chromatin elasticity is modeled by a Kratky-Porod model~\cite{Kratky:1949un},  a freely jointed chain of $N$ beads of diameter $a_0=1$, interconnected by tethers, whose length is constrained by  $a_0\le l a_0 \le \sqrt{3}a_0$ to ensure self-avoidance~\cite{Leibler:1987eq}, and is described by the energy,
\be 
E =\frac{\kappa}{2} \sum_{i=1}^{N} \left(1- {\bf t}_{i} \cdot {\bf t}_{i+1}\right) \,l_{i},
\label{kratskyporod}
\ee 
where 
$\kappa$ is the bending rigidity (in units of k$_{B}$T). The tangent vector, ${\bf t}_{i}$ is the unit vector from the $i^{th}$ bead to its nearest neighbor $i-1$ and  $l_{i}$ is the length of the vector, in units of $a_{0}$, from the bead at $i-1$ to $i+1$. The effective rigidity $\kappa$ of the coarse-grained chromatin is around $120$ \kbt{}~\cite{dekker:2013hi}, higher than the bare DNA  due to large steric interactions and the tight winding around histones.  We model confinement by a constraint on the   enclosed volume $V$ (in 3d), the Lagrange multiplier fixing this constraint is the pressure, $\Delta p$.

The mechano-chemical reactions discussed above are implemented in the DMC by adding and removing beads followed by re-tethering the polymer -- addition of a bead is equivalent to the ``forward'' reaction \ce{Chr  ->[k_{+}] Chr$^+$}, while removal of a bead is equivalent to the ``backward'' reaction \ce{Chr  ->[k_{-}] Chr$^-$}.
 
We summarize the DMC  moves : (i) Choose a particle and displace  it  within a small box of size $x$ around its old position, keeping the connectivity the same. The displacement is accepted using the Metropolis scheme~\cite{Frenkel:2001},  with the probability for acceptance given by, $P_{\rm acc}= \mbox{min} (1, e^{-\beta \Delta E})$, where  $\beta=1/$ \kbt{} is the inverse temperature and $\Delta E$ is the change in  the  energy through Eq.\,(\ref{kratskyporod}). We define a DMC time step (MCS) as $N$ attempts to displace the particles.  (ii) Remove or add a particle with rates that depend on the local curvature and projected density. These active moves are attempted $\epsilon_{\pm}$ times every MCS and is followed by  reconnection.
We briefly describe the active moves to set notation, for details, ({\it Supplementary Information, Secs.\,S2(A-C)}). A bead is added (removed)  to (from) a randomly chosen location on the polymer with probability,
\bea
&P_{+}=\frac{1}{ \left\{1+\exp\left[- \beta  \mu_{+} H_{+} \right] \right\} \left\{1+\exp\left[\gamma (N-\overline{N})\right]\right\} }\, , & \label{eqn:probadd} \\
&P_{-}=\frac{1}{ \left\{1+\exp \left[- \beta  \mu_{-} H_{-} \right] \right\} \left\{1+\exp\left[-\gamma (N-\overline{N})\right] \right\} } \,. \label{eqn:probrem} &
\eea
where  $\mu_{+}$ and $\mu_{-}$ are, respectively, the energy cost to add and remove from regions of unit positive curvature on the polymer, and could be thought of as a chemical potential. 
$H_{+}$ and $H_{-}$ are the curvatures that determine the chemical potentials for addition and removal of beads at the chosen location ({\it Supplementary Information, Sec.\,S1}) .
%$H_{+}$ and $H_{-}$ are the local mean curvatures to add and remove at the chosen location 
 
The parameter $\gamma$ controls the fluctuation of the number of particles about the expected value  $\overline{N}$, and is chosen to be high $\gamma = 0.1$, so as to get rms fluctuations less than 10\% ({\it Supplementary Information, Fig.\,S8}).

The addition and removal of monomeric units on the polymer  is accepted subject to fulfillment of the tether and self avoidance constraints. It is easy to see that these transition probabilities do not obey the detailed balance condition, 
\be  P(N)P_{+}(N\rightarrow N+1) \neq P(N+1)P_{-}(N+1\rightarrow N) , 
\ee when $\mu_{+}\neq \mu_{-}$, reflecting the non-equilibrium feature of the process.  Here, $P(N)$ is the probability to chose a vertex from a system containing $N$ vertices. The transition probabilities shown in eqns. \eqref{eqn:probadd} and \eqref{eqn:probrem} are purely non-equilibrium and detailed balance is satisfied when $\mu_{+}=\mu_{-}=0$.
%and $\gamma=0$. 

\section{Results}
We implement the DMC in both 2 and 3 dimensions; as discussed earlier the former corresponds to the conformations explored by chromatin when strong anchored to a two dimensional substrate. In any case, we will use the 2d model to show that it reproduces well known results in appropriate limits.
\subsection{Nonequilibrium phase diagram : active control of loop-on-loops}
\begin{figure*}[!h]
\centering
\includegraphics[width=15cm,clip]{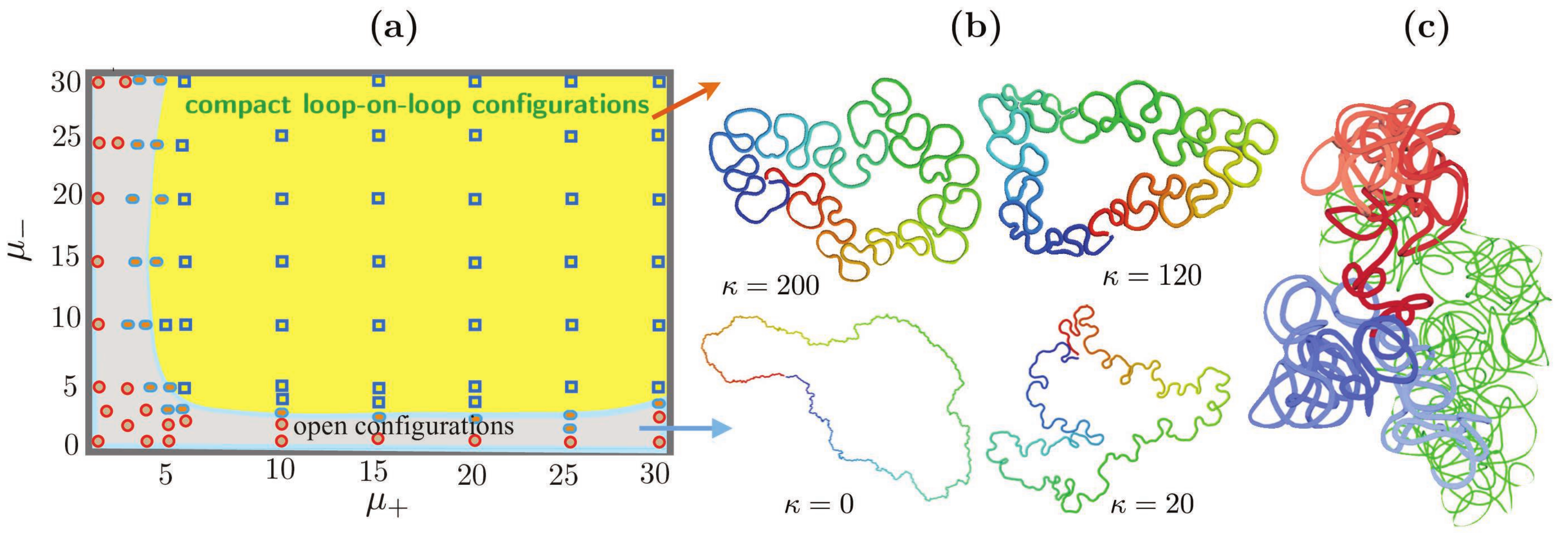}
\caption{\label{fig:phasedia}{\bf Steady state shapes of an active semiflexible polymer.} (a) Phase diagram of steady state conformations of an active ring polymer in 2d in the space of curvature-dependent active rates of adding (removing) beads, showing a transition from open to compact loop-on-loop conformations. Here $\overline{N}=2500$ and $\kappa=120$ \kbt{}  (b) Fixing the curvature-dependent active rates at $\mu_{+}=\mu_{-}=20$ \kbt{}, the 2d active ring polymer shows a transition from an open to a compact loop-on-loop conformation, upon increasing the bending rigidity $\kappa$. The short length scale fluctuations get ironed out on increasing activity.   (c) Looped or fibril-like conformations of a 3d active polymer with $\overline{N}=2500$, $\kappa=120$ \kbt{} and $\mu_{+}=\mu_{-}=20$ \kbt{}. Polymer segments are colored based on their position along the contour -- blue  and red beads respectively represent the start and end positions. The coloring on the active polymer conformations indicate that the colors are spatially segregated and do not intermingle, as seen in fractal globule models for chromatin organization~\cite{LiebermanAiden:2009jz,Mirny:2011cl}. }
\end{figure*}

We first check that in the absence of addition and removal of beads or when the removal and addition maintains detailed balance,  $\mu_{+}=\mu_{-}=0$, we recover the equilibrium phase diagram of a closed semiflexible polymer in  2-dimensions in the canonical and the grand canonical ensembles, respectively ({\it Supplementary Information, Sec.\,S2(D-H)}). We then restore full  activity and vary the parameters, $\mu_{+}, \mu_{-}$, $\kappa$ and $\Delta p$ to obtain a phase diagram in $d=2$, Fig.~\ref{fig:phasedia}(a). We perform various tests to confirm that these structures are indeed true steady states of an  active polymer ({\it Supplementary Information, Sec.\,S2(I-J)}).
When the bending rigidity $\kappa$ is small such that the polymer size is larger than the persistence length, it behaves like a self-avoiding random walker (SAW), showing fluctuations at the smallest scales. As we increase $\kappa$, these small scale fluctuations get ironed out -- at $\kappa=120$, the active semi-flexible polymer shows {\it loop-on-loop} structures (Fig.\,\ref{fig:phasedia}(b)). The final conformation of the polymer does not depends on its initial state. One of our predictions is that the looped structure is formed only beyond a critical bending rigidity $\kappa_c$ for a fixed value of $N$ and activity. Further, for fixed $\kappa$ and activity, looped structures are obtained for large enough $N$.  We find it difficult to arrive at a systematic classification of shapes and hence a phase diagram in $d=3$. Even so it is apparent from Fig.~\ref{fig:phasedia}(c), that the 3d conformations show a tendency to form correlated fibril-like structures \cite{wolynes:2015}. We now turn to a study of the  statistical properties  of the active polymer at steady state in both $d=2$ and $3$. 
\begin{figure*}[!h]
\centering
\includegraphics[width=6.0in,clip]{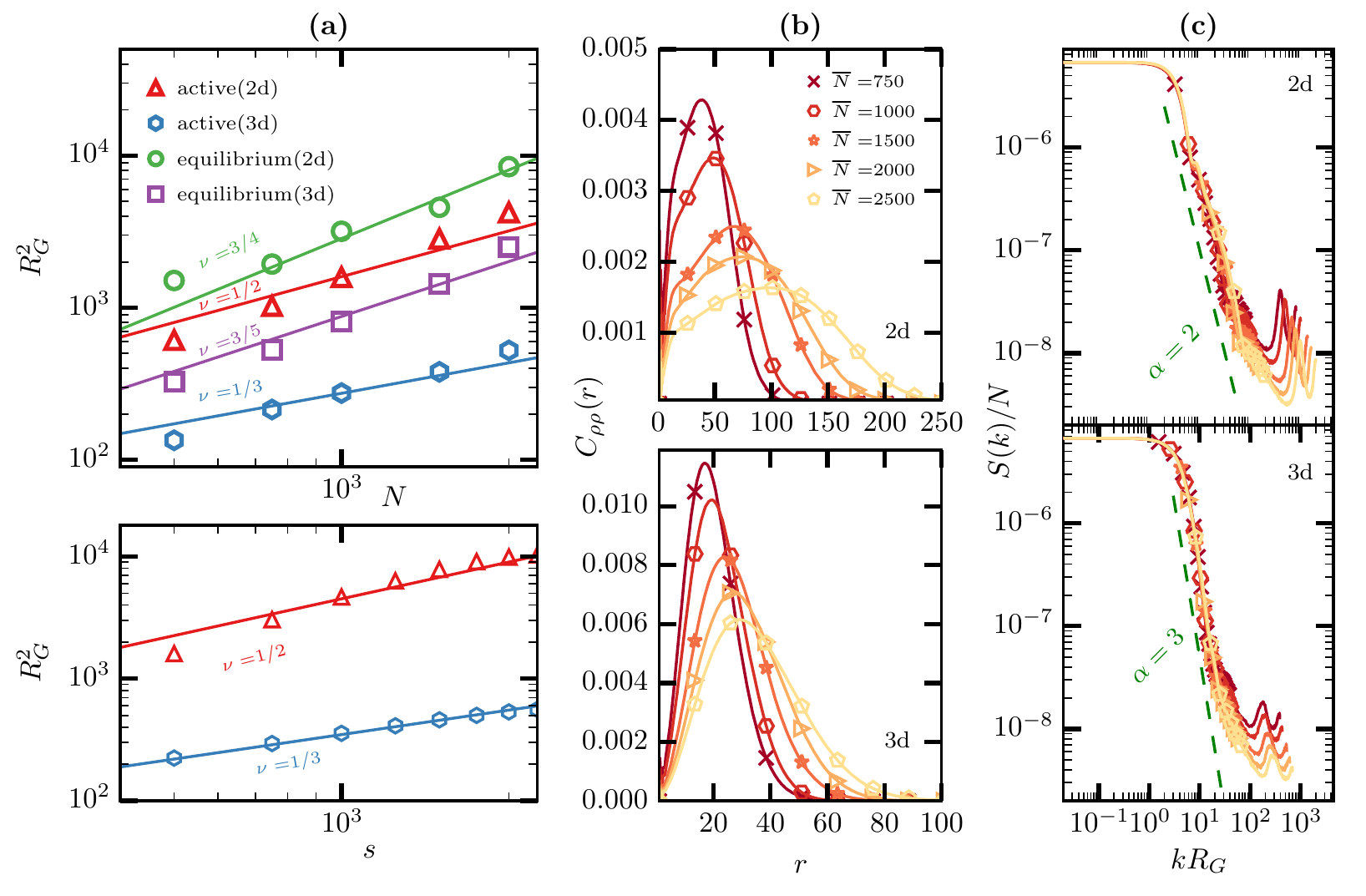}
\caption{\label{fig:rg-denden} {\bf Statistics of an active polymer in 2d and 3d.} (a)  ({\it top}) Scaling $R_{G}^{2}\sim N^{2\nu}$ of radius of gyration with size $N$.
We first verify that the  equilibrium ring and linear polymer in 2d and 3d is consistent with self-avoiding random walk exponents, $\nu = 3/4$ (2d) and  $\nu = 3/5$ (3d). The active polymer is seen to  be {\it compact} with  $\nu \approx 1/2$ (2d) and $\nu \approx 1/3$ (3d).  The rest of the parameters are  $\kappa=120$ \kbt{}, $\mu_{+}=\mu_{-}=20$ \kbt{}, and $\epsilon_{\pm}=0.1$ \nmcs{}. ({\it bottom})  Indeed every subchain of a sufficiently long active polymer (exceeding $\xi_p$, the persistence length) is also compact, showing $R_{G}^{2}(s) \sim s^{2\nu}$, for $s > s_{min}$, with $\nu \approx 1/2$ (2d) and $\nu \approx 1/3$ (3d) ($\overline{N}=2500$). (b) Density correlation function $C_{\rho \rho}$ of an active polymer in 2d ({\it top}) and 3d ({\it bottom}) with spatial distance $r$ also indicates a more compact structure than the corresponding equilibrium polymer ({\it Supplementary Information, Fig.\,S13}) -- data shown for $\overline{N}$ in the range $750$ to $2500$, with $\kappa=120$ \kbt{} and $\epsilon_{\pm}=0.1$ \nmcs{}. (c) Corresponding scaled structure factor $S(k)/N$ as a function of the scaled wavenumber $x=kR_{G}$ in 2d ({\it top}) and 3d ({\it bottom}), showing a power law decay $S(x) \sim x^{-\alpha}$ with exponents $\alpha \approx 2$ and $\alpha \approx 3$, respectively.  }
\end{figure*}
\subsection{Active polymer is compact, space-filling and self-similar}
We analyze the scaling behaviour of the radius of gyration ($R_{G}^{2}$) for an active semiflexible polymer, with linear and ring topologies, with respect to polymer size $N$. We find that the active semiflexible polymer is {\it compact} with its radius of gyration scaling as $R_{G}^{2}\sim N^{2\nu}$, with $\nu \approx 1/2$ in 2d  and  $\nu \approx 1/3$ in 3d (Fig.\,\ref{fig:rg-denden}(a)). As control, we verify that we recover the scaling behaviour of an equilibrium self-avoiding polymer when the activity is switched off (Fig.\,\ref{fig:rg-denden}(a)). We note that the statistics obtained in the case of ring polymers is especially clean (data not shown). We also verify that every subchain of sufficiently long polymer is collapsed, i.e., a fit to $R_{G}^{2}(s) \sim s^{2\nu}$, for $s > s_{min}$, shows that $\nu \approx 1/2$ and $\approx 1/3$, in $d=2$ and $3$, respectively (Fig.\,\ref{fig:rg-denden}(a)). Thus the polymer starts becoming compact at a value of $s$ significantly smaller than the confinement scale (or the size of the polymer), a finding that appears consistent with recent FISH experiments ~\cite{MateosLangerak:2009go}. This is likely due to the natural tendency of the active polymer to form multiple loops.
\subsection{Density correlations and structure factor}
We compute the monomer density correlation $C_{\rho\rho}(r)=N^{-1}\langle \rho(0)\rho(r)\rangle$ of the active, self-avoiding, semi-flexible polymer in 2d and 3d, for both linear and ring topologies (Fig.\,\ref{fig:rg-denden}(b)). For the equilibrium polymer, the value of $r$ at which the density correlator peaks is roughly set by the persistence length, $\xi_p$. In contrast, for the active polymer, the peak of the density correlation is at smaller $r$ and is set by activity. This scale  $\xi_{a}$ is related to the loop-size that we discuss  next. The structure factor $S(k)$, the Fourier transform of the density correlation, also shows distinct features of activity. We focus on the form of the structure factor at intermediate wave vector scales, $2\pi/R_G < k < 2 \pi/b$. For the active polymer in 2d, we find that $S(k) \sim k^{-2}$; whereas for the 3d active polymer, $S(k) \sim k^{-3}$ (Fig.\,\ref{fig:rg-denden}(c) and {\it Supplementary Information, Fig.\,S14}). This is consistent with the predicted behaviour of the fractal or crumpled polymer~\cite{Halverson:2014hg}.
\subsection{Statistics of Looping} 
The most significant result from the recent Hi-C experiments ~\cite{LiebermanAiden:2009jz} is that the probability of contact and loop formation of  human interphase chromatin is distinct from an equilibrium polymer. Here we show that the active polymer in 3d shows a contact statistics that has the same scaling behaviour as that reported in~\cite{LiebermanAiden:2009jz}. The contact probability density $P(r(s))$, is defined as the probability to find two coarse grained beads, separated along the polymer contour by $s$, to be at a spatial (geometric) distance $r(s)$ ({\it Supplementary Information, Fig.\,S15}).  The heat map for $P(r(s))$ computed for both equilibrium and active polymers are shown in the {\it Supplementary Information, Fig.\,S16}. The equilibrium polymer with finite $\kappa$ shows a band like structure implying the lack of any long range contact, with a spread that increases with lowering $\kappa$.  In spite of its diffuse nature we observe reasonable contact density only for small values of the contour length $s<100a_{0}$ which indicates the absence of any order promoting long range contacts. In contrast, the contact probability density for an active polymer shows finite contact density for beads separated by contour length $s<400a_{0}$ indicating the presence of compactness and order.  
 
\begin{figure*}[!h]
\centering
\includegraphics[width=15cm,clip]{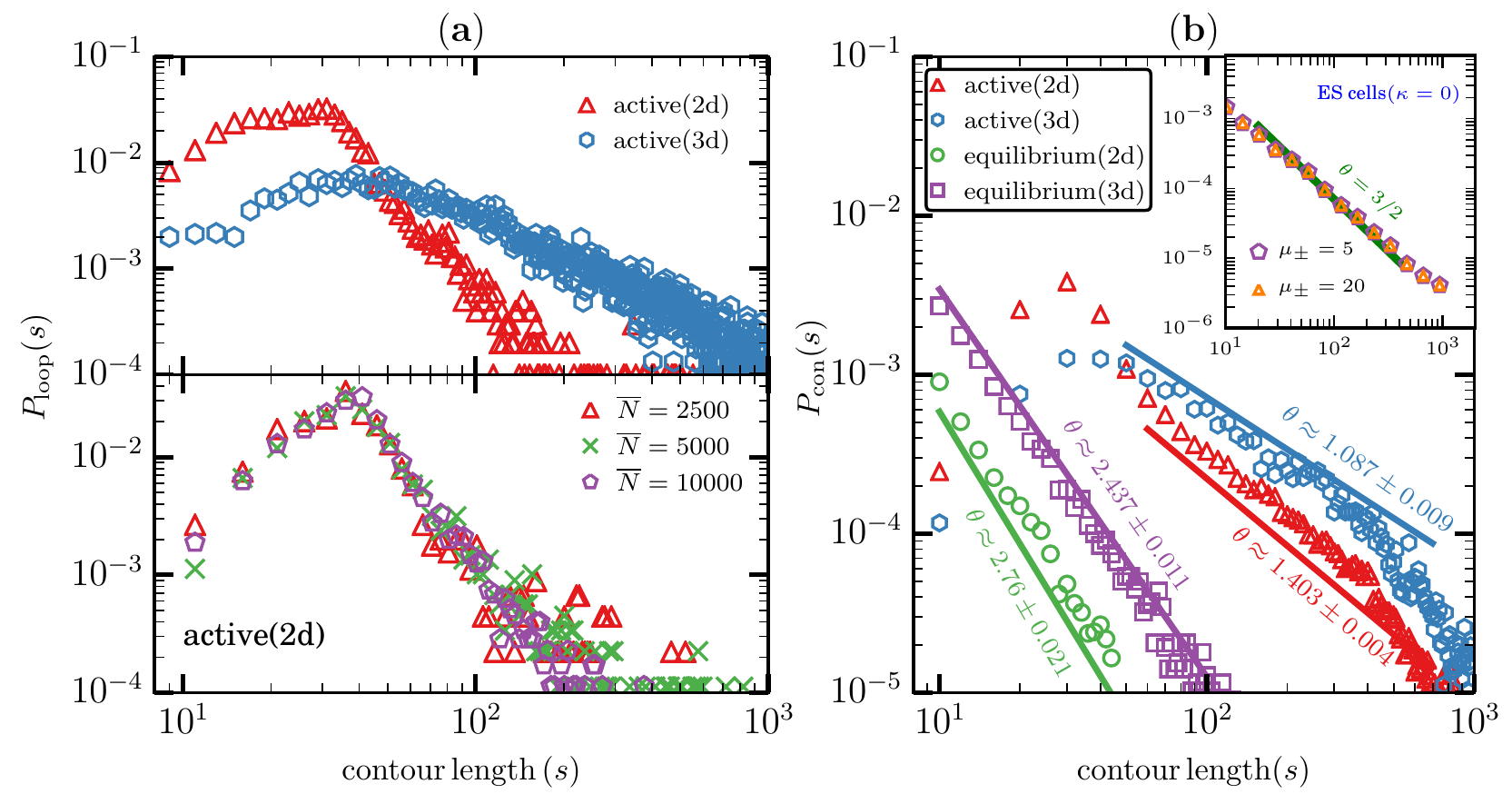}
\caption{\label{fig:loopsize} {\bf Looping and contact probabilities of an active polymer.} (a) ({\it top}) Loop size distribution as a function of $s$, the contour length of the active polymer, with $\overline{N}=2500$ in 2d and 3d. ({\it bottom}) Loop size distribution for active polymers in 2d for  $\overline{N}=2500$, $5000$, and $10000$. (b) Contact probability  of an active ring polymer  scales as $P_{\rm con}(s) \sim s^{-\theta}$, with $\theta \approx 1.403\pm 0.004$ (active 2d),  $\theta \approx 1.087\pm 0.009$ (active 3d), $\theta \approx 2.76\pm 0.021$ (equilibrium 2d) and $\theta \approx 2.437\pm 0.011$ (equilibrium 3d). Observed scaling for the active polymer is consistent with $\theta =  3/2$ ($d=2$) and $\theta = 1$ ($d=3$), expected of a fractal polymer. Here, $\overline{N}=2500$, $\kappa=120$ \kbt{}, $\Delta p=0$ and $\mu_{+}=\mu_{-}=20$ \kbt{}. Inset shows scaling of the contact probability for the 3d active polymer when $\kappa=0$ \kbt{}, which we take to represent chromatin configurations in ES cells, with $\theta\approx 3/2$, consistent with \cite{wolynes:2015}. 
}
\end{figure*}

 We first determine the loop size distribution $P_{\rm loop}(s)$ of the active polymer, defined as the probability for two coarse grained beads, separated by $s$, to form a loop like structure (Fig.~\ref{fig:loopsize}(a), and also see {\it Supplementary Information, Fig.\,S18} for details).  Our simulations of the active semiflexible polymer in 2d, seem to suggest that there is a typical loop size, which is independent of the size of the polymer, and dependent on activity. In 3d, however, the loop size distribution is very broad at the peak and overlaps with the region where the contact probability $P_{\rm con}$ (to be defined below) shows an
exponential cutoff, making it difficult to ascribe a typical loop size. We note that the steady state loops of the active polymer are dynamic, although its movement is suppressed by topological constraints and viscosity. 
 
We next study the scaling behaviour of the contact probability $P_{\rm con}(s)=P(r(s))\,\,\forall \,(a_{0}<r(s)<\sqrt{3}a_{0})$ for the active semiflexible self-avoiding polymer in 2 and 3 dimensions (Fig.~\ref{fig:loopsize}(b)). As control, we first display the contact probability of the equilibrium self-avoiding polymer. In 2d, the contact probability exhibits a power-law decay with  $P_{\rm con}(s) \sim s^{-\theta}$, where $\theta = 2.76\pm0.02$ at large enough $s$. The value of this contact exponent is in excellent agreement with the exact theoretical value of $\theta=2.68$ obtained using conformal invariance~\cite{Duplantier:1987vw}. Note that the probability of contact is very nearly zero, for scales smaller than the persistence length, $\xi_{p}$. In 3d, the contact probability for the equilibrium polymer also decays as a power law with an exponent $\theta = 2.437\pm 0.001$. Recall that the contact exponent for a Gaussian polymer is $\theta=d/2$ in $d$-dimensions.
For an active semiflexible self-avoiding polymer, however, one obtains loops at scales smaller than $\xi_p$, and it is only below scales corresponding to  $\xi_a$ (set by activity) that  the contact probability is zero -- at larger scales $s \gg \xi_a$ we find that $P_{\rm con}(s) \sim s^{-\theta}$, with $\theta \approx 3/2$ ($d=2$) and $\theta \approx 1$ ($d=3$) (Fig.~\ref{fig:loopsize}(b)). The scaling of the contact probability in 3d is in excellent agreement with the Hi-C data of  interphase chromatin in mammalian cells~\cite{LiebermanAiden:2009jz}. 
On the other hand, the contact exponent has been reported to be $\theta \approx 3/2$ in human ES -- to explain this, we recognise that the chromatin in ES cells are transcriptionally very active and show large fluctuations and lower percentage of compact heterochromatin regions. This we argue corresponds to setting $\kappa \approx 0$ in our active chromatin model -- our simulations then show that $\theta \approx 3/2$ (inset, Fig.~\ref{fig:loopsize}(b)). As we argued, the contact statistics of chromatin in yeast cells should more closely resemble the statistics of the active polymer in $d=2$.

\subsection{Note on knots}
It has been argued ~\cite{Halverson:2014hg} that the space-filling, fractal polymer is a solution of the optimal folding of a polymer that is confined and constrained to be in a knot free state. Unfortunately our algorithm to generate the active dynamics (the dual description) uses  non-topological moves of monomer addition and removal and polymer reconnection. The resulting knottedness, which we verify using the knots server (\url{http://knots.mit.edu}~\cite{Virnau:2006jt}), is entirely an artifact of our algorithm. We expect that we can make local remedial  moves on this steady state conformation to anneal the knots  without changing the statistical features of the active polymer, such as $R_G^2$ and $P_{\rm con}$. Notwithstanding, the topological constraints and the smooth dynamics of the active polymer would ensure a knot free steady state when the initial condition is knot free, both for a closed polymer and a long enough linear polymer~\cite{Halverson:2014hg}.

\subsection{Segmented activity}
Being a compact globule even at scales smaller than the confinement scale, there is topological repulsion between two non-concatenated active polymer rings. However, pressure from confinement of a bunch of nonconcatenated ring polymers can result in a degree of intermingling, which can be modulated by having an intervening stretch of inactive chromatin of length $l$ (``euchromatin'', see {\it Supplementary Information, Fig.\,S19}). This lack of intermingling and entanglement of active polymers is consistent with recent observations~\cite{Halverson:2014hg}.

\subsection{Turning off activity}
One way to check if activity is the driver of in-vivo chromatin folding at large scales, is to switch off the effects of activity, either by ATP depletion or by perturbing the activity of chromatin remodeling proteins (as in~\cite{Hameed:2012ej}), and monitor the ensuing dynamical changes. We carry out this protocol in our simulations, i.e., starting from the steady state of the active polymer, we monitor the dynamics of conformational changes upon switching off activity.  We find that the polymer very quickly evolves to a more swollen conformation, with the conformational statistics of an equilibrium self-avoiding polymer ({\it Supplementary Information, Fig.\,S3}). This would suggest that the nuclear volume should increase upon reducing or switching off activity, consistent with the findings reported in \cite{Talwar:2013bh}.

\section{Discussion}
We have seen how the statistics of the steady state conformations of a semiflexible self-avoiding polymer subject to active stresses replicates the observed statistical features of interphase chromatin at a megabase scale as reported by the recent 
Hi-C data~\cite{LiebermanAiden:2009jz}. This might suggest that active stress generation together with topological constraints are important ingredients in driving chromatin folding at large scales.  We find that hierarchical looping appears as  a natural outcome of activity and semiflexibility, and does not  necessarily require the engagement of specialized bridging or linker proteins, although these play important roles in stabilizing specific loop regions and regulating their dynamics and territorial segregation.

It is encouraging to us that this active chromatin model shows reasonably good agreement with the Hi-C results on the statistics of packing and contact. In addition, it makes many predictions such as, (i) the scaling of the structure factor, $S(k) \sim k^{-3}$, (ii) the broadness of the probability distribution of loop sizes $P_{\rm loop}(s)$, and (iii) the low stiffness of active chromatin in ES cells. 

Based on our numerical evidence and the simulation results of~\cite{LiebermanAiden:2009jz}, we make two further observations regarding the conformations of the active polymer  -- (i) in two dimensions, we find that the statistics of the active semi-flexible self-avoiding polymer is the same as that of the space-filling Peano (or Hilbert) curves, and (ii) in three dimensions, we find that the statistics of the active semi-flexible self-avoiding polymer is the same as that of the space-filling  {\it interdigitating} Peano curve. Indeed in 2d, the similarity between the typical loop-on-loop conformation of the active polymer (Fig.\,\ref{fig:phasedia}(b)) and Hilbert curves is quite striking.

Space-filling Peano or Hilbert curves often come up in spatial ordering of data structures or
 in the design of optimal data storage in computer science. Typical database applications require multi-attribute indexing, which involves  mapping points from a multidimensional space into one dimension in a way that preserves spatial ``locality'' {\it as much as possible}.  Locality implies points that are close together in multi-dimensional attribute space are also close together in the one-dimensional space~\cite{Faloutsos:1989wn}. This kind of mapping is important in assigning physical storage to minimize access effort~\cite{Jagadish:1990bu}.
It appears from many studies that by ordering data along space-filling curves, such as Peano or Hilbert curves, one best achieves this goal.

Given that the statistical properties of the conformation of interphase chromatin is the same as space filling Peano or Hilbert curves, it is tempting to conjecture that this folding architecture  optimizes information storage and access, by co-locating gene loci 
with shared transcription resources within loops. This appears consistent with very recent 
high resolution Hi-C studies, which studies the interplay between chromatin looping and gene-loci contacts \cite{dekker:2013hi}. 

Optimizing information processing cannot come for free, and it is appealing that in the model of active chromatin folding this is a consequence of  driving the system  out-of-equilibrium by 
energy consuming processes. It would be interesting to explore these ideas in studies combining such high resolution metrical studies with gene expression data.

\begin{acknowledgments}
NR thanks NCBS for hospitality. We thank M. Thattai for useful discussions and 
pointing us to Z-ordering and GV Shivashankar for comments on the manuscript.
\end{acknowledgments}

%\bibliographystyle{unsrtnat}
%\bibliographystyle{pnas}
%\bibliography{bibfile-31Dec14}

\pagebreak
\clearpage
{\Large \begin{center}Supplementary Information \end{center}}
\setcounter{equation}{0}
\setcounter{section}{0}
\setcounter{figure}{0}
\setcounter{page}{1}
\renewcommand{\thefigure}{S\arabic{figure}}
\renewcommand{\thesection}{S\arabic{section}}
\renewcommand{\theequation}{S\arabic{equation}}

\section{Energy Functional and Active Dynamics of a Confined Active Semiflexible Polymer}
\subsection{Effective free energy functional}
We assume that at equilibrium, the configurations of chromatin at large scales, are determined by polymer elasticity and 
its interaction with the multicomponent nucleoplasm, a highly viscous medium with an assortment of DNA-packaging proteins, such as histones. Taking a coarse grained view, the nuclear medium can be assumed to provide a mean-field  nuclear confining potential $U_{con}(r)$, which we can chose to be the harmonic oscillator potential or a volume constraint in the fixed pressure ensemble,  $\int d{\bf r} \, \rho U_{con}({\bf r}) \propto \Delta p V$ --- where the internal pressure, $\Delta p\equiv p_{in}-p_{out}$,  is  a result of  an outward pressure, arising from a combination of polymer entropy, steric interaction with the nucleoskeleton and coupling to the nuclear membrane via lamins, and an inward pressure mediated via non local attractive interaction between DNA-packaging proteins and possibly the confining pressure exerted by the cytoskeleton. 

The elastic interaction of the coarse-grained chromatin contains a bend and stretch elasticity (and a twist, which we will ignore), as
shown in Fig.\,1 (main text).
Thus in our coarse-grained description, the effective chromatin is {\it compressible}, and described by the following free-energy functional,
\begin{equation}
E  =  \int ds \left( \sigma + \frac{\kappa}{2}  \l H-H_0\r^{2} + \frac{k_s}{2} \l\epsilon-\epsilon_0\r^2
+ k_c \epsilon H \right) +\int dV \Delta p 
\label{free}
\end{equation}
where $H$ and $\epsilon$ are the local curvature and stretch deformations, respectively. 
The parameters, $\kappa$, $\sigma$, $H_{0}$, $k_s$, $\epsilon_0$ and $k_c$ are the bending modulus, tension, spontaneous curvature, stretch modulus, spontaneous strain and the bend-stretch coupling, respectively. In  general these parameters could be  heterogeneous since they depend on the local concentration of bound DNA-packaging proteins. Our Monte Carlo analysis uses a discrete version of the above functional as the Hamiltonian.

\clearpage
\newpage

\subsection{Curvature dependent active (un)binding kinetics}
The effective chromatin is driven out-of-equilibrium by the binding/unbinding of ATP-dependent chromatin-remodeling proteins (CRPs). 
We model this chemical (un)binding by a Michelis-Mentin (MM) kinetics, with reaction rates dependent on local DNA deformation, the curvature $H$ and stretch $\epsilon$ --- thus the CRPs can be either curvature or stretch sensors. If the unbound CRPs in solution form
a bath, then the local dynamics of the concentration of bound CRPs $\rho$ has the form

\be
\frac{d\rho}{dt} = \left\{ 
\begin{array}{l}
	k_+ f(H, \epsilon) - k_{-} \rho \\
	\,\,\,\,\,\,\,\,\,\,\,\,\,\,\, \mbox{or}
	\\
	\dfrac{k_1 \rho}{1+K_1 \rho} f(H, \epsilon) - k_{-} \rho\, , \\ \end{array} \right.
\label{chemreact} 
\ee
the latter depicting MM saturation kinetics, where $k_1$ and $K_1$  represent the usual constants in an MM reaction. The function $f$ describes the mechano-chemical coupling; note that in introducing mechano-chemical rates for binding alone, we have explicitly violated detailed balance, an essential ingredient in active chemical reaction kinetics.

We expect the form of $f$ with respect to deformations should generically be a constant for small deformation, then increase sharply before going to zero for sufficiently large deformations (as can be derived from a simple strain modified potential energy profile
characterizing the bound and unbound states). This picture of the mechano-coupling forms the basis for the form of the transition probabilities in the Dynamical Monte Carlo simulations.

\section{Dynamical Monte Carlo Simulations and Phase Diagram of the Active Polymer}

Our Dynamical Monte Carlo (DMC) approach works in the coarse grained  {\it dual representation} of the chromatin, as described in the main text (Fig.\,1). In this dual representation, the ``semiflexible polymer'' is treated as a connected set of beads, each of which represents a finite stretch of chromatin. The coarse-grained length and 
energy scales have been given in the main text. The dynamics of the polymer consists of the usual equilibrium polymeric moves and CRP-induced active moves corresponding to addition and removal of beads. The active addition of a bead represents a unit gain of a free stretch of chromatin length due the the CRP-induced release of histones. Likewise, the active removal of a bead represents a unit loss of a free stretch of chromatin length due the binding of histones onto that portion of the naked dsDNA. In our coarse grained representation, each bead corresponds  approximately to a $4$\,kbp stretch of dsDNA.

\subsection{Dynamical Monte Carlo moves} \label{sec:act-mcsmoves}
\begin{figure}[!h]
	\begin{center}
		\includegraphics[width=6.0in]{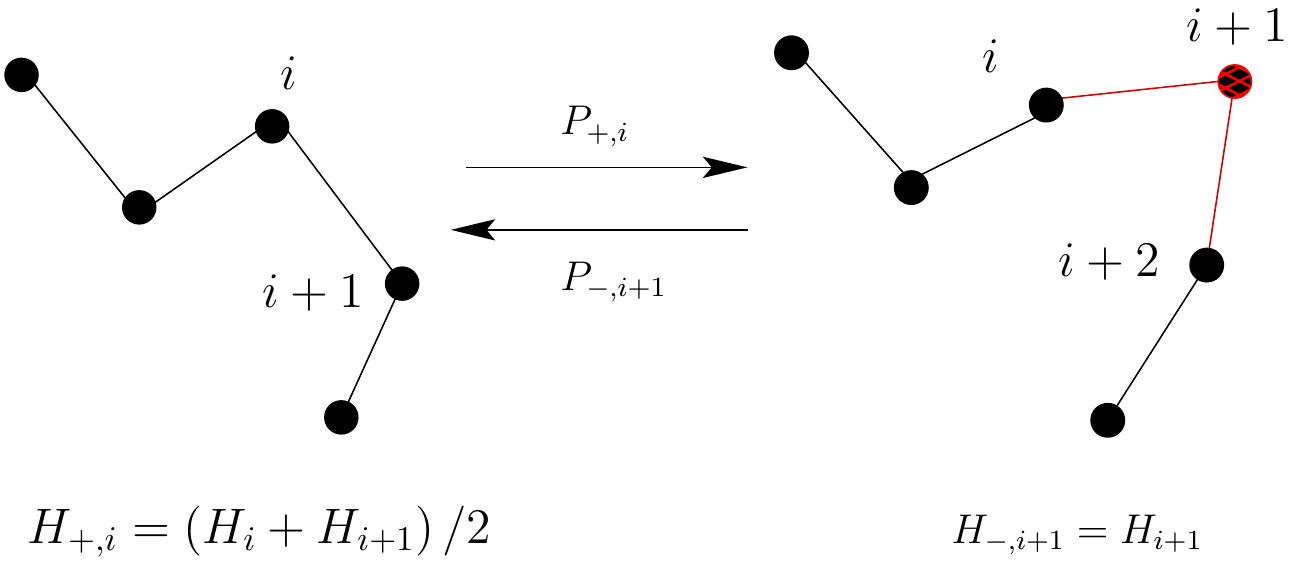}
		\caption{Scheme to add and remove a polymer bead from a segment of the polymer. (left) Conformation of a polymer with $N$ beads to which the $N+1^{\rm th}$ bead is added, at a location between the $i$ and $i+1^{\rm th}$ bead, with a probability $P_{+,i}$ and (right) conformation of a polymer segment with $N+1$ beads from which the $i+1^{\rm th}$ bead is removed with a probability $P_{-,i+1}$. The mean curvatures  that determine the addition and removal rates are also shown. }
		
		\label{fig:active-model}
	\end{center}
\end{figure}

All accessible states in the configuration space of the inactive and active polymers are sampled by a set of Monte Carlo moves described  below:
\begin{enumerate}[(a)]
	\item  {\it Bead move}: A randomly chosen polymer bead is displaced within a square (cube) of side $2\sigma$, whose center coincides with the center of the bead and is oriented with its edges parallel to the cartesian axes.
	
	This displacement changes the state of the polymer from $\left[\{\vec{X}\},N\right] \rightarrow \left[\{\vec{X}^{'}\},N\right]$ and its energy from $E \rightarrow E^{'}$, where $\{\vec{X}\}$ is the set of all position vectors of the polymer beads. 
	The move is accepted using the Metropolis scheme with probability,
	\begin{equation}
	P_{\rm dis}={\rm min}\left \{1, \exp(-\beta (E^{'}-E)) \right \},
	\end{equation}
	which obeys detailed balance.
	
	\item{\it Addition and Removal moves}: The active process of histone-release (histone-wrapping) results in the increase (decrease) of a stretch of polymer, which
	in our model results in the addition (removal) of beads at a randomly chosen location, as shown in Fig. \ref{fig:active-model}. 
	
\end{enumerate}

\begin{enumerate}[(i)]
	\item \textsf{Addition of bead: } A bead is placed inside a square (cube) centered on the tether connecting the randomly chosen bead $i$ and its neighbor $i+1$. If the self avoidance constraint is satisfied the beads are reconnected such that bead $i+1$ is now $i+2$. This addition move, at bead $i$, is attempted with a curvature dependent probability,
	\begin{equation}
	\label{eq:padd}
	P_{+,i}=\frac{1}{\left\{1+\exp({- \beta  \mu_{+} H_{+,i})}\right\} \left \{1+\exp(\gamma (N-\overline{N})) \right\}}\,\,,\,\,\,\,  
	\end{equation}
	where $H_{+,i}=(H_i+H_{i+1})/2$ is the mean of the curvatures at beads $i$ and $i+1$, which are  $H_{i}$ and $H_{i+1}$ respectively. $\overline{N}$ denotes the average desired size of the polymer and $\mu_{+}$ is the energy cost to add a bead on a tether with unit mean curvature, that can be identifiable with a chemical potential for curvature field. $\gamma$ is a dimensionless parameter that controls the width of number (length) fluctuations of  the polymer around the imposed  steady state value $\overline{N}$ .

	\item \textsf{Removal of bead:} As illustrated in Fig. \ref{fig:active-model}, a chosen polymer bead  $i+1$ is removed when $a_{0}<\vert\vec{X}(i)-\vec{X}(i+2)\vert<\sqrt{3}a_{0}$. The move is attempted with a probability,
	\begin{equation}
	\label{eq:prem}
	P_{-,i+1}=\frac{1}{ \left \{1+\exp({- \beta  \mu_{-} H_{-,i+1})} \right \} \left \{1+\exp (-\gamma (N-\overline{N})) \right \} } \,\,.
	\end{equation}
	$H_{-,i+1}=H_{i+1}$ is the curvature at $i+1$ and $\mu_{-}$ is the energy cost to scission off a bead from a segment of unit curvature.
\end{enumerate}
Unlike the displacement move, whose acceptance depends on the energy of initial and final states and obeys detailed balance, the active  addition/removal moves do not depend on the change in the energy of the polymer. Instead these moves are attempted with probabilities defined in eqns. \eqref{eq:padd} and \eqref{eq:prem}  as long as 
the self avoidance constraint is met. In the next subsection we show that this active move explicitly violates detailed balance, as it should.

Each Monte Carlo step thus contains $N$ attempts to perform a random walk and $\epsilon_{\pm}$ attempts to add and remove beads from the polymer.

Note that for a polymer in three dimensions the curvature measure is chosen to be the local unsigned mean curvature  computed as $H_{i}=\sqrt{(1-\hat{t}_i \cdot \hat{t}_{i+1})}$. For a polymer confined to the two dimensional  $xy$ plane, one can also use the signed mean curvature which is  estimated as $H_{i}=\sqrt{(1-\hat{t}_i \cdot \hat{t}_{i+1})}\,\, {\rm sgn}\left (\hat{z} \cdot (\hat{t}_i \times \hat{t}_{i+1}) \right )$.

%%%
%%%%
\subsection{Active transition probabilities and the Kolmogorov loop condition}{\label{sec:kolmogorov}
	We now explicitly show that the form of transition probabilities for active moves (eqns.~\eqref{eq:padd} and ~\eqref{eq:prem}) do not obey detailed balance, by demonstrating a violation of the
	Kolmogorov loop condition. The Kolmogorov loop condition states  that for every loop in state space, the product of the transition probabilities in one direction is equal to the product taken in the reverse direction. Our task is therefore to construct a loop where this condition is violated.
	\begin{figure}[!h]
		\centering
		\includegraphics[width=15cm,clip]{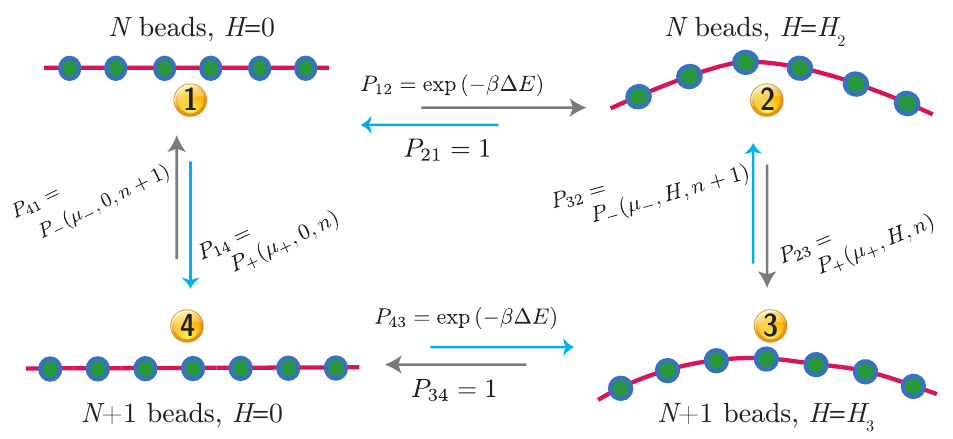}
		\caption{\label{fig:kolmogorov} A Kolmogorov loop diagram illustrating the transition probabilities between four distinct states of the active polymer. States 1 and 2 have $N$ beads each, while states  3 and 4 each have $N+1$ beads.}
	\end{figure}
	
	Consider four distinct states of the active polymer, marked as 1, 2, 3, and 4 in Fig.~\ref{fig:kolmogorov}. These different states are characterized by the difference in the number of coarse grained beads and in the curvature of the backbone :
	\begin{enumerate}[]
		\item {\sf STATE 1}: $N$ beads with curvature at every bead $H=0$
		\item {\sf STATE 2}: $N$ beads with curvature at every bead $H = H_{2}$
		\item {\sf STATE 3}: $N+1$ beads with curvature at every bead $H = H_{3}$
		\item {\sf STATE 4}: $N+1$ beads with curvature at every bead $H=0$
	\end{enumerate}
	
	Let $P_{\alpha \beta}$ denote the transition rate (which could be either equilibrium or active)  between any two given states $\alpha$ and $\beta$. For the example considered in Fig.~\ref{fig:kolmogorov}, bead number preserving transitions ($P_{12}$, $P_{21}$, $P_{34}$, and $P_{43}$) are equilibrium transitions, while those transitions that lead to a change in the bead number ($P_{23}$, $P_{32}$, $P_{41}$, and $P_{14}$) are active transitions. If $\Delta E$ is the change in energy during an equilibrium transition due to a change in the backbone curvature, then the equilibrium transition rate is given by  min(1,$\exp(-\beta \Delta E)$). It can be shown that  $P_{12}=P_{43}=\exp(-\beta \Delta E)$ and $P_{21}=P_{34}=1$.
	
	Similarly, the active transition rates can be computed using eqns.~\eqref{eq:padd} and ~\eqref{eq:prem} (also see eqns. (4) and (5) in the main manuscript). Consider the transition from state 2 ($N$ beads and curvature $H_{2}$) to state 3 ($N+1$ beads and curvature $H_{3}$). If the number deviation is $N-\overline{N}=n$, then using equation (4) we can show that,
	\begin{equation}
	P_{23}=\frac{1}{\left\{1+\exp(-\beta \mu_{+}H_{2} \right\}\left\{1+\exp(\gamma n) \right\}}.
	\end{equation}
	Similarly, the other transition rates are given by,
	\begin{eqnarray}
	P_{32}& = & \dfrac{1}{\left\{1+\exp(-\beta \mu_{-}H_{3} \right\}\left\{1+\exp(-\gamma (n+1)) \right\}},   \\ 
	& & \nonumber \\
	P_{14}& = & \dfrac{1}{2\left\{1+\exp(\gamma n) \right\}},  \\
	& & \nonumber \\
	P_{41}& = & \dfrac{1}{2\left\{1+\exp(-\gamma (n+1)) \right\}}. 
	\end{eqnarray}
	
In systems where microscopic reversibility is obeyed, the Kolmogrorov loop condition states that the clockwise and counter-clockwise transition probabilities are related by,  
\begin{equation}
P_{12} P_{23}P_{34}P_{41} - P_{14}P_{43}P_{32}P_{21} = 0\, .
\end{equation}
The clockwise cyclic transition probabilities ($1 \rightarrow  4$),  $P_{12}P_{23}P_{34}P_{41}=\left(1+\exp(-\beta \mu_{+}H_{2}) \right)^{-1}$, while the counter  clockwise cyclic transitions ($4 \rightarrow  1$), $P_{14}P_{43}P_{32}P_{21}=\left(1+\exp(-\beta \mu_{-}H_{3}) \right)^{-1}$. The two probabilities are different  when $\mu_{+}, \mu_{-} \neq 0$, since generically the curvature of the two states, $H_{2}$ and $H_{3}$, are different. This violation of the Kolmogorov loop  condition explicitly demonstrates the violation of detailed balance.
	
\subsection{Efficient implementation of the addition and removal moves}
The addition and removal moves are extremely constrained by the self avoidance condition imposed on the polymer beads. We find that a majority of the randomly selected positions to add/remove a bead violates this condition and hence a majority of the attempted moves are rejected leading to inefficient sampling of the conformations of the active polymer. 

The active conformations can be sampled with a much higher efficiency if the active moves, treated as rare events, are sampled using enhanced sampling techniques. A number of such methods exist for rare event sampling in equilibrium Monte Carlo~\cite{Frenkel:2001}.  In the current context, we employ the following procedure : if while adding(removing) a bead to(from) a polymer configuration, there is a potential violation of the self avoidance(tethering) constraint, then instead of rejecting that move, we locally expand(compress) the polymer configuration to accommodate the addition(removal) of the bead.  An expansion move is shown in Fig.~\ref{fig:polymer-expansion}, where an attempt is made to add a bead, as described in Sec.~\ref{sec:act-mcsmoves}, at a randomly chosen position $\vec{r}_{*}$, located between beads $i$ and $i+1$. Such moves assume faster relaxation of the polymer through tangential moves. 
\begin{figure}[!h]
\centering
\includegraphics[width=15cm]{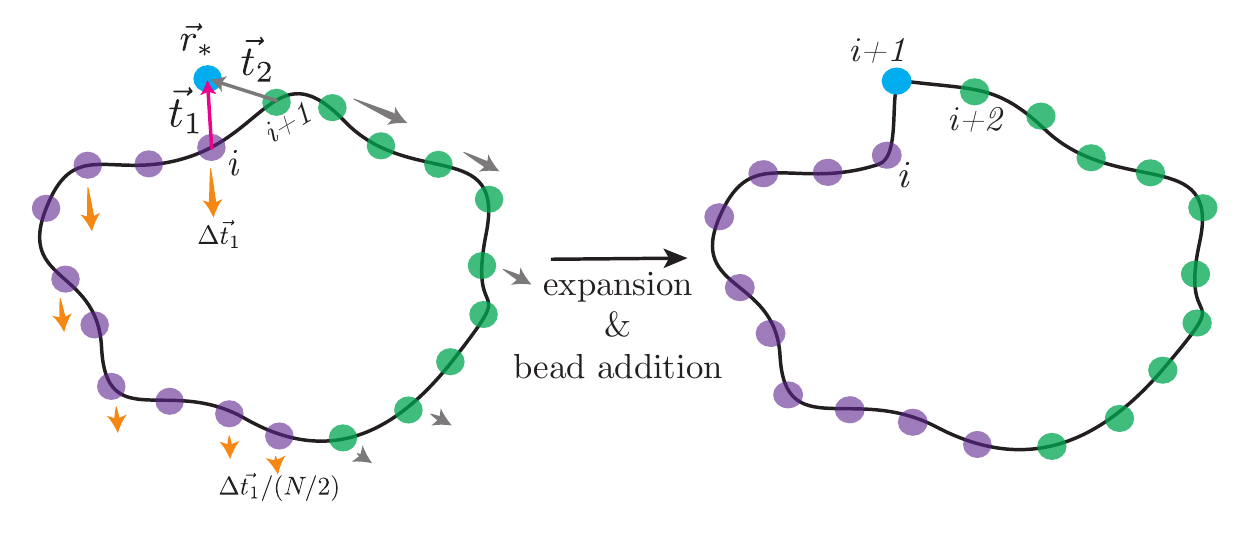}
\caption{\label{fig:polymer-expansion} An illustration of polymer expansion employed in the active addition of beads for a two dimensional ring polymer whose initial configuration is shown in the left panel. The additional bead is inserted at a location $\vec{r}_{*}$ whose relative position with respect to beads $i$ and $i+1$ are $\vec{t}_{1}$ and $\vec{t}_{2}$ respectively. The polymer is expanded by individually displacing each of the polymer beads,  with the displacement at a given bead being proportional to its separation with respect to bead $i$ along the polymer contour.	The tapered arrows represent the direction and magnitude of the displacement at each of the beads in the polymer chain. The configuration post expansion is shown in the right panel.}
	\end{figure}
	
If $\vec{r}_{i}$ and $\vec{r}_{i+1}$ are the positions of  beads $i$ and $i+1$, then $\vec{t}_{1}=\vec{r}_{*}-\vec{r}_{i}$ and $\vec{t}_{2}=\vec{r}_{*}-\vec{r}_{i+1}$ are the relative vectors of the new bead with respect to $i$ and $i+1$. The additional bead overlaps with $i$ or $i+1$ when $\sqrt{3}a_{0}<|\vec{t}_{1}|<a_{0}$ or $\sqrt{3}a_{0}<|\vec{t}_{2}|<a_{0}$ respectively. We overcome such a  violation by displacing the bead(s) violating the self-avoidance constraint by $\Delta \vec{t}_{m}=(|\vec{t}_{m}|-\langle l \rangle)\hat{t}_{m}$, where $m=1$ or $2$.   Here $\langle l \rangle$ is the average distance between any two consecutive beads in an equilibrium polymer --- we compute this quantity from independent simulations and based on these studies we set  $\langle l \rangle = 1.3a_{0}$. The local displacement of beads $i$ and $i+1$ may lead to large local deformations in the polymer chain that can affect its relaxation dynamics. To prevent any such conformational artifacts we also individually displace all the other beads in the polymer. In our procedure, the displacement of a bead is chosen to be proportional to its separation along the polymer chain either with respect to bead $i$ or $i+1$. The displacement vector for the individual beads is calculated as follows :
\begin{enumerate}[(a)]
\item divide the polymer chain with $N$ beads into two equal sized segments, each with $N/2$ beads, as shown in Fig.~\ref{fig:polymer-expansion}
\item the first segment comprises of beads from $i$ to $(i-(N-2)/2)$ (subject to periodic boundary conditions) and these beads are indexed from $m=1...(N-2)/2$, such that the local index of bead $i$ is $(N-2)/2$
\item the second segment contains all the beads between $i+1$ and $(i+1+(N-2)/2)$, with respective indices $m=1,....,(N-2)/2$, such that the local index of bead $i+1$ is1 $(N-2)/2$
\end{enumerate}
	
\noindent The displacement of the individual beads in the first segment is given by
\begin{equation*}
\Delta {\vec t}_{m,1}=2m \frac{\Delta \vec{t}_{1}}{(N-2)} \quad {\rm with} \quad  \Delta {\vec t}_{1}=(|\vec{t}_{1}|-\langle l \rangle)\hat{t}_{1},
\end{equation*}
and the displacement vector of the beads in the second segment is given by
\begin{equation*}
\Delta {\vec t}_{m,2}=2m \frac{\Delta {\vec t}_{2}}{(N-2)} \quad {\rm with} \quad  \Delta {\vec t}_{2}=(|\vec{t}_{2}|-\langle l \rangle)\hat{t}_{2}.
\end{equation*}
	
The expanded polymer configuration is accepted as the new configuration, along with the newly added $(N+1)^{\rm th}$ bead, if all the bead positions satisfy the self avoidance constraint. A schematic of the final polymer conformation with $N+1$ beads is shown in the right panel of Fig.~\ref{fig:polymer-expansion}.

\subsection{Scaling properties of the equilibrium ring polymer in two dimensions} \label{sec:lsf}

We first validate our MC algorithm in well known limits. With this in mind, we study the equilibrium properties of the {\it inactive} ring polymer in 2 dimensions and test our results against the well known results of the Leibler-Singh-Fisher (LSF) model~\cite{Leibler:1987eq}. 
	
The observed conformations of the 2d polymer match very well with the shapes predicted in~\cite{Leibler:1987eq} - a floppy polymer ($\kappa=0$) displays branched polymer configuration at deflating pressures ($\Delta p \leq 0$) and an extended shape for inflating pressures ($\Delta p>0$).  We also check the finite size scaling of geometric quantities, such as the enclosed area and the squared radius of gyration~\cite{Leibler:1987eq}
\begin{eqnarray}
R^2_G (N,\Delta p) &\approx & N^{2 \nu} \Lambda(\overline{p} N^{\phi \nu}), \\
& & \nonumber \\
A(N,\Delta p) &\approx &N^{2 \nu} \Omega(\overline{p} N^{\phi \nu}),  
\end{eqnarray}
with the scaling variable given by the scaled osmotic pressure difference  $\overline{p}=\Delta p a^2/k_BT$. The values of the scaling exponents are $\phi \sim 2.13$, $2\nu \sim 1.51$. The form of the scaling functions $\Lambda$ and $\Omega$ (Fig. \ref{fig:lsf-scale}) is entirely consistent with the results of the Leibler-Singh-Fisher (LSF) model~\cite{Leibler:1987eq}.}

\begin{figure}[!h]
\centering
\includegraphics[width=12.5cm]{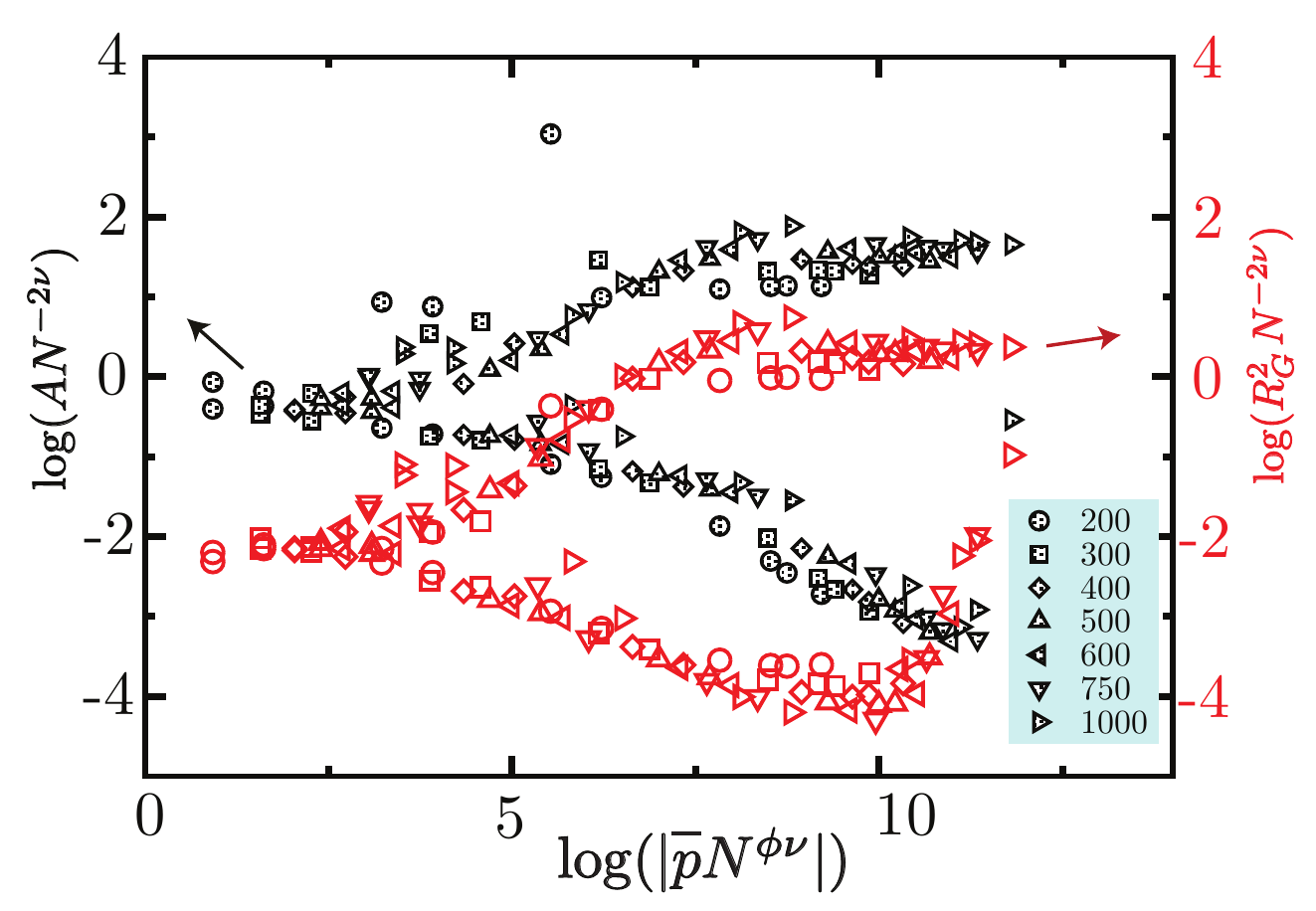}
\caption{\label{fig:lsf-scale} {Finite-size scaling collapse of the enclosed area (filled symbols) and $R^2_G$ (unfilled) for polymer sizes in the range of $N=100$ to $1000$. The scaling behaviour of the equilibrium polymer in 2 dimensions is entirely consistent with the LSF predictions~\cite{Leibler:1987eq}.} }
\end{figure}

\subsection{Conformations of the equilibrium polymer}
\begin{figure}[!h]
\centering
\includegraphics[width=15cm]{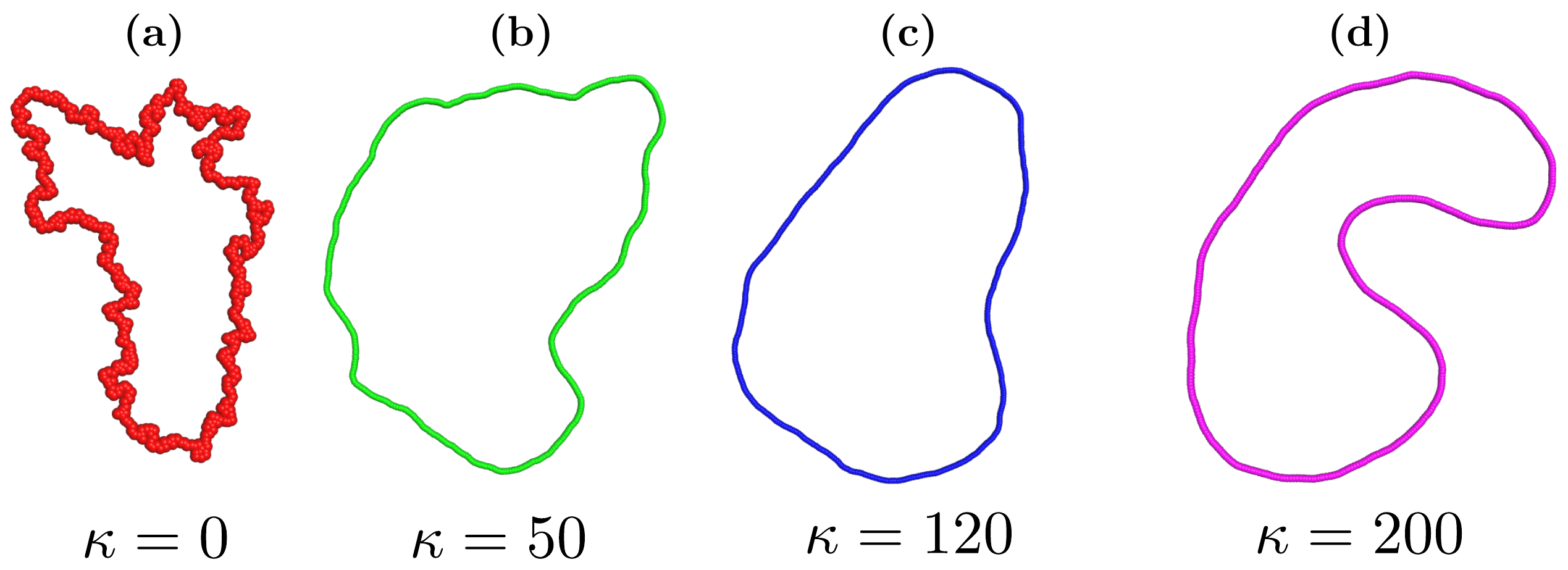}
\caption{\label{fig:equm-conf-zeropr} Conformations of an equilibrium ring polymer with $N=2493$ beads as a function of the bending rigidity, with the osmotic pressure difference $\Delta p=0$.}
\end{figure}

\begin{figure}[!h]
\centering
\includegraphics[width=12.5cm]{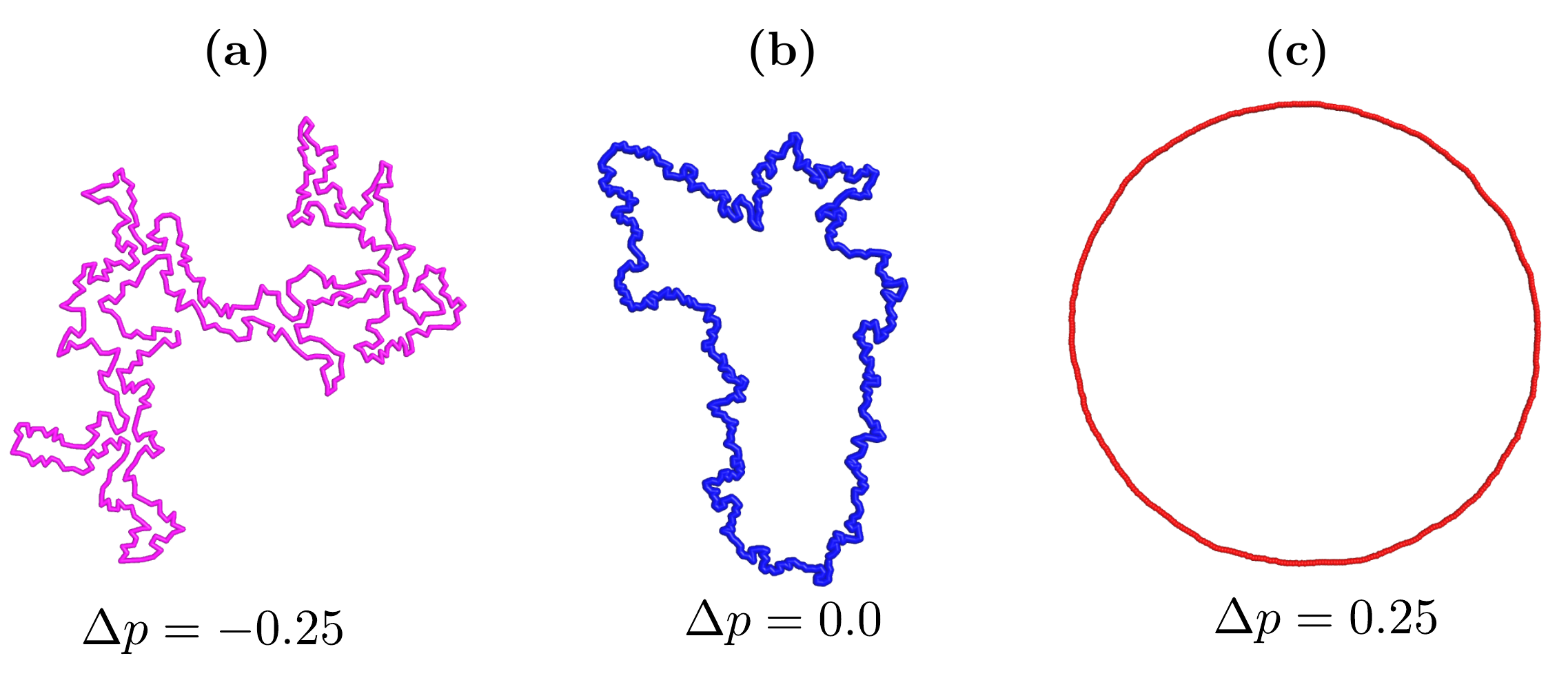}
\caption{\label{fig:equm-0-conf-funcpr} Conformations of an equilibrium ring polymer, with $N=500$ beads and $\kappa=0$, as a function of osmotic pressure difference $\Delta p$.  }
\end{figure}

\begin{figure}[!h]
\centering
\includegraphics[width=12.5cm]{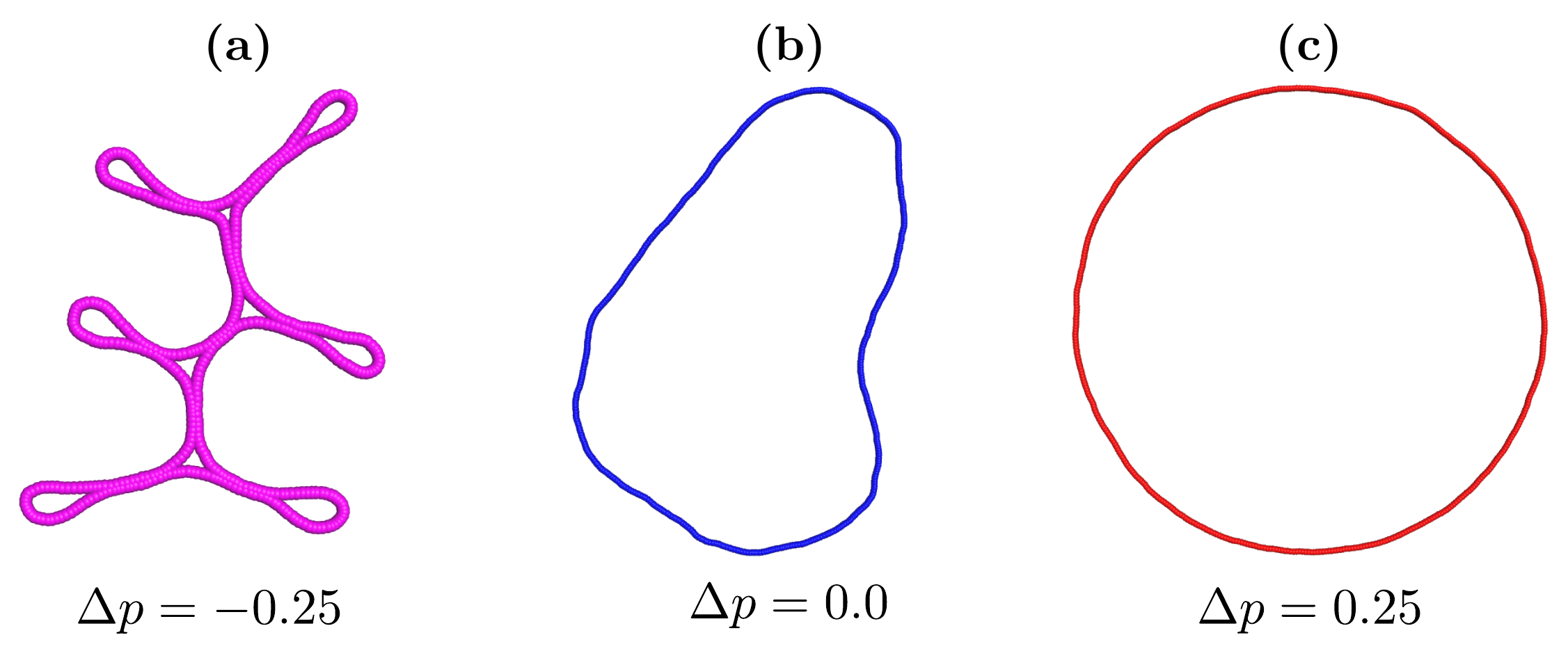}
\caption{\label{fig:equm-120-conf-funcpr} Conformations of an equilibrium ring polymer, with $N=500$ beads and $\kappa=120$ \kbt{}, as a function of osmotic pressure difference $\Delta p$. }
\end{figure}

The equilibrium conformations of a ring polymer in two dimensions as a function of the bending rigidity $\kappa$ at $\Delta p=0$, is shown in Fig.~\ref{fig:equm-conf-zeropr}. When $\kappa=0$ (flexible polymer),  the configuration of the polymer is governed by entropic contributions leading to a rough structure at all length scales, as is shown in Fig.~\ref{fig:equm-conf-zeropr}(a). In the semi-flexible, and stiff limit the conformational landscape of the polymer is governed by the bending energy and hence the polymer shapes are correlated upto the persistence length of the polymer, as shown in Figs. \ref{fig:equm-conf-zeropr}(b) and (c).
The osmotic pressure difference ($\Delta p$) is another key factor that impacts the equilibrium shape of a ring polymer. Figs.~\ref{fig:equm-0-conf-funcpr} and \ref{fig:equm-120-conf-funcpr} show the configurations of a two dimensional ring polymer as a function of $\Delta p$. When $\Delta p <0$, the polymer experiences a deflating pressure and forms  packed configurations that minimize the enclosed area. Snapshots of ring polymers subjected to a deflating pressure ($\Delta p=-0.25$) are shown in Figs. \ref{fig:equm-0-conf-funcpr}(a) and \ref{fig:equm-120-conf-funcpr}(a), respectively for $\kappa=0$ and $120$ \kbt{}.

Similarly, the presence of an inflating pressure maximizes the area enclosed by the ring polymer. Hence the polymer displays the inflated configurations as shown in Figs. \ref{fig:equm-0-conf-funcpr}(c) and \ref{fig:equm-120-conf-funcpr}(c), respectively for $\kappa=0$ and $120$ \kbt{} . The shapes of the ring polymer for both values of $\kappa$ when $\Delta p=0$ is shown in Figs. \ref{fig:equm-0-conf-funcpr}(b) and \ref{fig:equm-120-conf-funcpr}(b) for comparison.

\subsection{Time series of $N$ due to addition/removal of beads}

For the simulations of the two dimensional active polymer we need to go to system sizes which are an order of magnitude larger than what we have done for the 
equilibrium polymer. Thus the system sizes under study involve upto ${N}\sim 1000, 2000$ and even $2500$ beads. The fluctuations in the number of beads as a result of the active addition-removal is controlled by the parameter $\gamma$ given in eqns. \eqref{eq:padd} and \eqref{eq:prem}. 

\begin{figure}[!h]
\centering
\includegraphics[width=16.5cm]{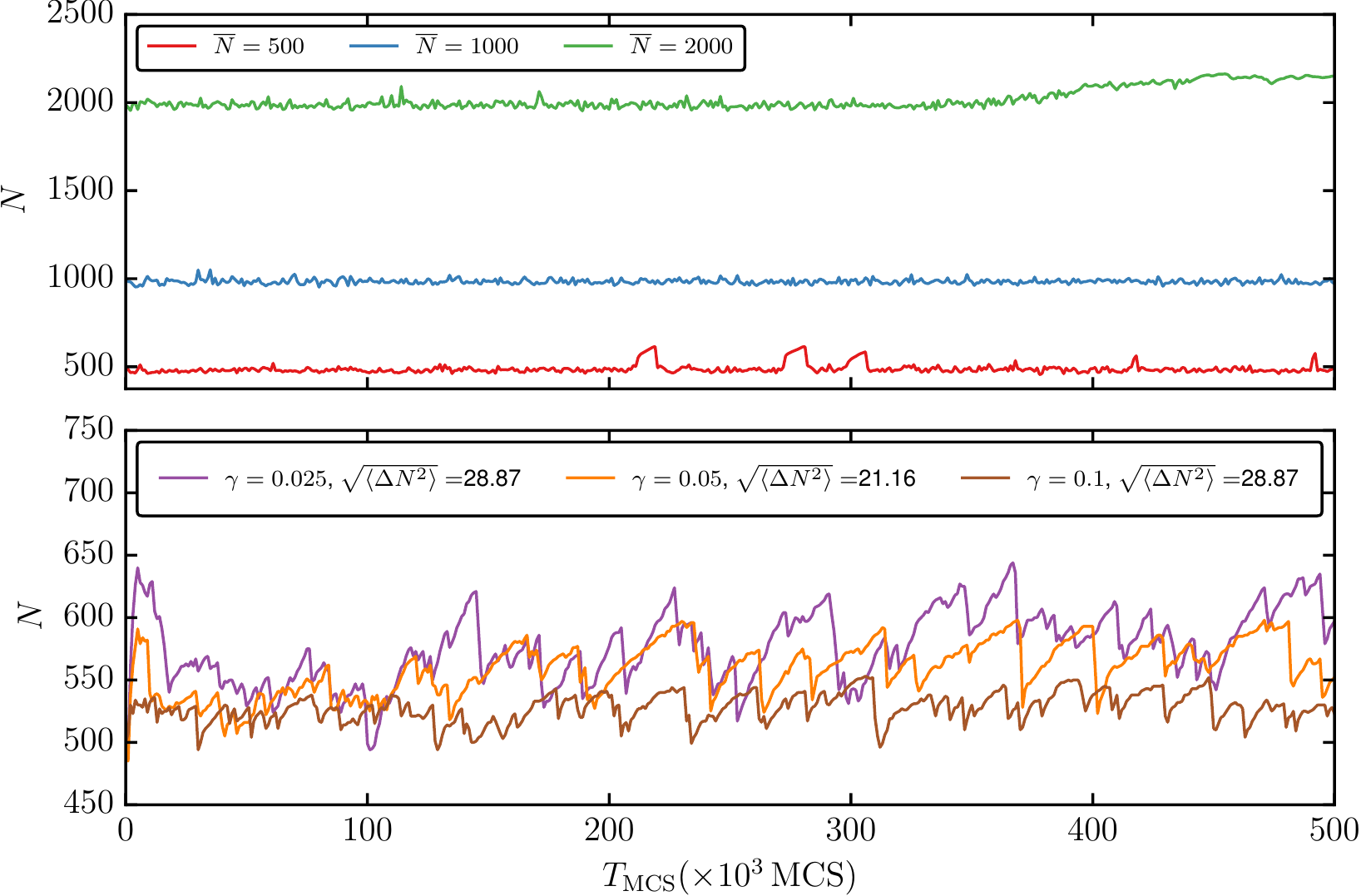}
\caption{\label{fig:timeseries-N} {\it (top panel)} Number of beads in the 2d active polymer as a function of Monte Carlo time for three different values of $\overline{N}=500$, $1000$, and $2000$ at $\gamma=0.05$. {\it (bottom panel)} Time series of $N$  for three different values of $\gamma=0.025,\,0.05$ and $0.1$ with a fixed  $\overline{N}=500$. The width of the number fluctuations measured by the standard deviation $\sqrt{\langle \Delta N^{2} \rangle}$ decreases with increasing value of $\gamma$ (as indicated in the legend). All data correspond to a 2d active ring polymer with $\kappa=120$ \kbt{}, $\epsilon_{\pm}=0.1$ \nmcs{} and $\mu_{+}=\mu_{-}=20$ \kbt{}.}
\end{figure}

The top panel in Fig.~\ref{fig:timeseries-N} shows the time series (in MCS times) of the number of beads in an active polymer for three different values of the expected number, $\overline{N}=500$, $1000$, and $2000$, at a fixed value of the parameter $\gamma=0.05$ defined in eqn.~\eqref{eq:prem}. 
The effect of varying $\gamma$ can be seen in the lower panel of Fig.~\ref{fig:timeseries-N} where we have shown the time series of the number of beads in a 2d active ring polymer for three different values of $\gamma=0.025,\,0.05$ and $0.1$ for a fixed value of $\overline{N}=500$. Also shown are the corresponding values of the standard deviation in the number of beads $\sqrt{\langle \Delta N^{2} \rangle}$. Consistent with eqn.~\eqref{eq:prem} we find that the number fluctuations  decrease with increasing $\gamma$. For all the studies reported in the main manuscript we have used a fixed value of $\gamma=0.05$.

\subsection{Conformations of the active polymer as a function of $\kappa$}
\begin{figure}[!h]
\centering
\includegraphics[width=12.5cm]{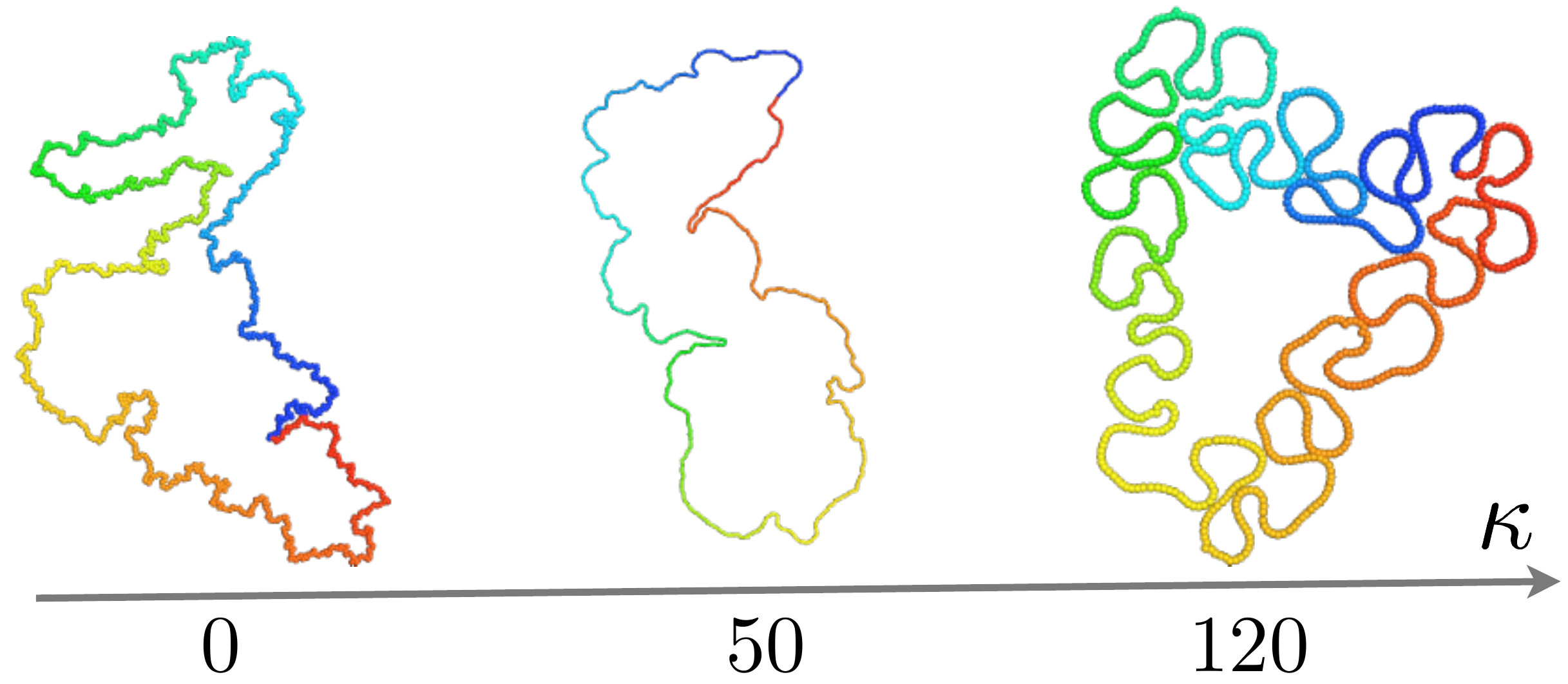}
\caption{\label{fig:conf-func-kappa} Configurations of a 2d active ring polymer as a function of $\kappa$, for fixed values of $\mu_{+}$=$\mu_{-}$=20 \kbt{}, $\Delta p=0$, and $\epsilon_{\pm}=0.1$ \nmcs.}
\end{figure}

The shape of the active polymer is a strong function of the bending rigidity $\kappa$, as shown in Fig. \ref{fig:conf-func-kappa} for three different values of the bending rigidity.
Increasing the rigidity $\kappa$ irons out the small wavelength fluctuations and generates a loop-on-loop structure.

\subsection{Conformations of the active polymer as a function of $\mu_{+}$ and $\mu_{-}$}
\begin{figure}[!h]
\centering
\includegraphics[width=12.5cm]{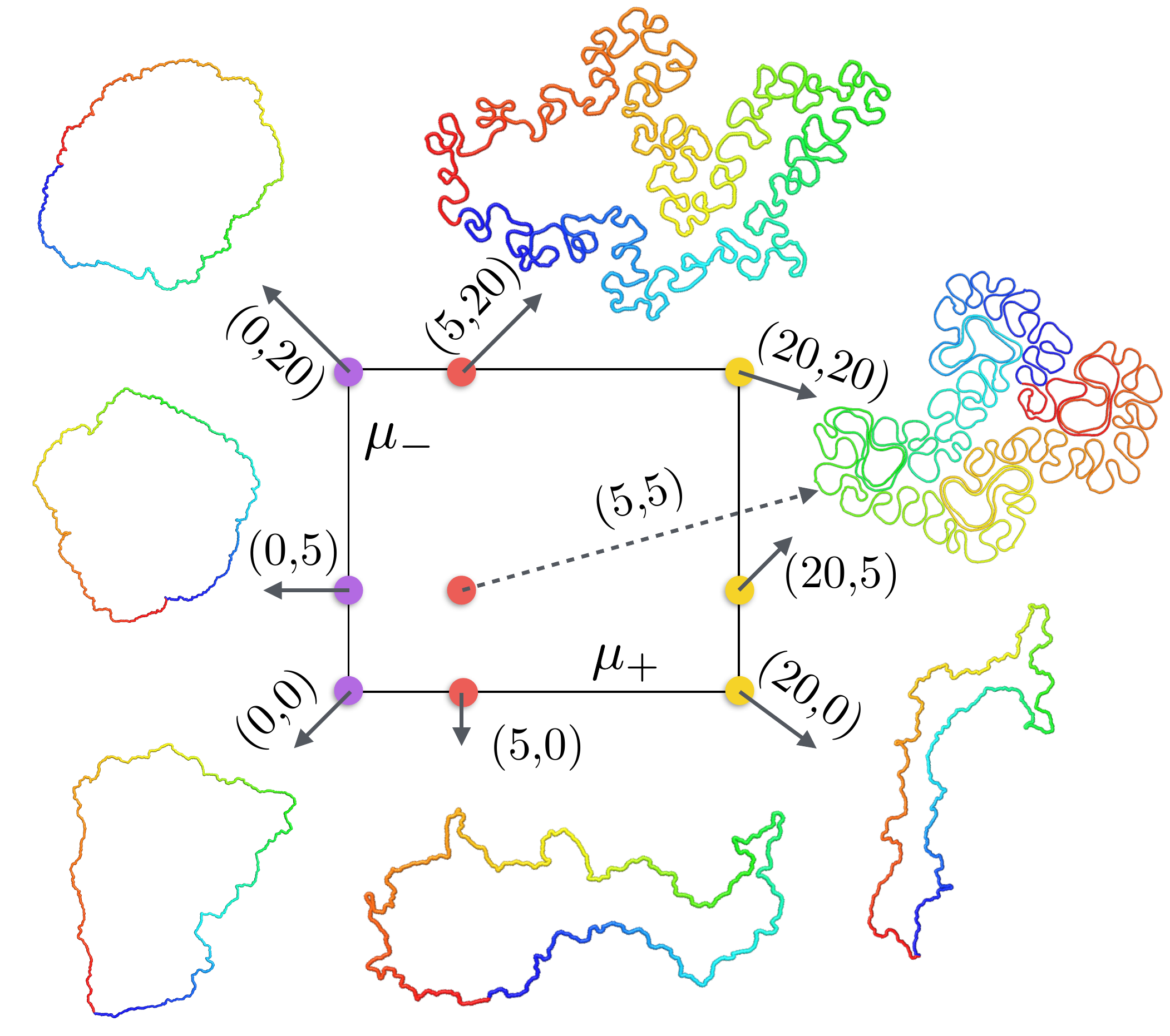}
\caption{\label{fig:conf-func-madd-mrem} Configurations of a two dimensional, ring, active polymer as a function of $\mu_{+}$ and $\mu_{-}$, for $\kappa=120$ \kbt{}, $\Delta p=0$, and $\epsilon_{\pm}=0.1$ \nmcs.}
\end{figure}

The conformations of the active polymer is also a function of the nonequilibrium  chemical potentials $\mu_{+}$ and $\mu_{-}$. Fig. \ref{fig:conf-func-madd-mrem} shows the steady state configurations exhibited by an active, two dimensional, ring polymer with $\overline{N}=2500$, $\kappa=120$ \kbt{}, $\Delta p=0$ and activity rate $\epsilon_{\pm}=0.1$ \nmcs  \,\,for various values of $\mu_{+}$ and $\mu_{-}$. The active polymer undergoes a change from extended to compact shapes when the chemical potential exceeds a critical value. The conformation phase diagram for the parameters listed above, is shown in the main text.

\subsection{Time series of $R_{G}^{2}$}
\begin{figure}[!h]
\centering
\includegraphics[width=15cm]{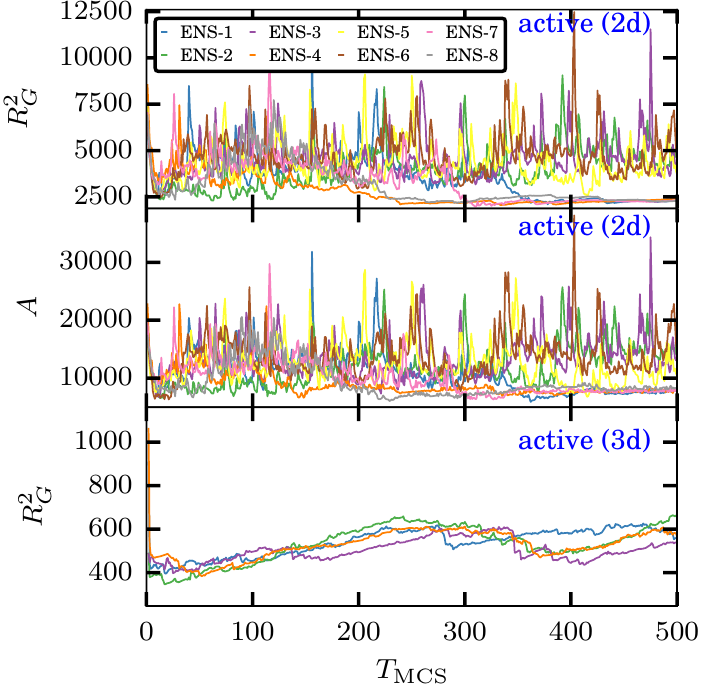}
\caption{\label{fig:rg2-area-vsT} {\it (top and center panels)} $R_{G}^{2}$ and area $A$ as a function of simulation time for eight different ensembles of a 2d active ring polymer with $\overline{N}=2000$, $\kappa=120$ \kbt{}, $\mu_{+}=\mu_{-}=20$ \kbt{} and $\epsilon_{\pm}=0.1$ \nmcs. {\it (lower panel)} $R_{G}^{2}$ as a function of simulation time for an active polymer in 3d.}
\end{figure}

The looped structures of an active polymer continuously expand and shrink in response to the addition and removal of particles. This behavior is evidenced in the large fluctuations in geometric measures like the radius of gyration and the enclosed area. The instantaneous values of $R_{G}^{2}$ and $A$ for eight different ensembles of a two dimensional active polymer are shown in the top and center panels of Fig.~\ref{fig:rg2-area-vsT}\textemdash each of these ensembles differ in their starting configuration and initial random seed but all have the same set of simulation parameters fixed to be $\overline{N}=2000$, $\kappa=120$ \kbt{}, $\mu_{+}=\mu_{-}=20$ \kbt{} and $\epsilon=0.1$ \nmcs. The value of $R_{G}^{2}$ for an active polymer in three dimensions is shown alongside in the bottom panel of Fig.~\ref{fig:rg2-area-vsT}.

The fluctuations in the polymer conformations are noticeably larger in two dimensions compared to their three dimensional counterpart. The statistics of these fluctuations are governed by the curvature dependent chemical potentials $\mu_{+}$ and $\mu_{-}$. However the number of degrees of freedom to add and remove particles is limited in the case of two dimensions and as a result the active polymer undergoes a rapid disassembly in order to generate more degrees of freedom for the particles added in the next Monte Carlo step.

\subsection{Looped configurations are characteristic steady state of an active polymer}
\begin{figure}[!h]
\centering
\includegraphics[width=12.5cm]{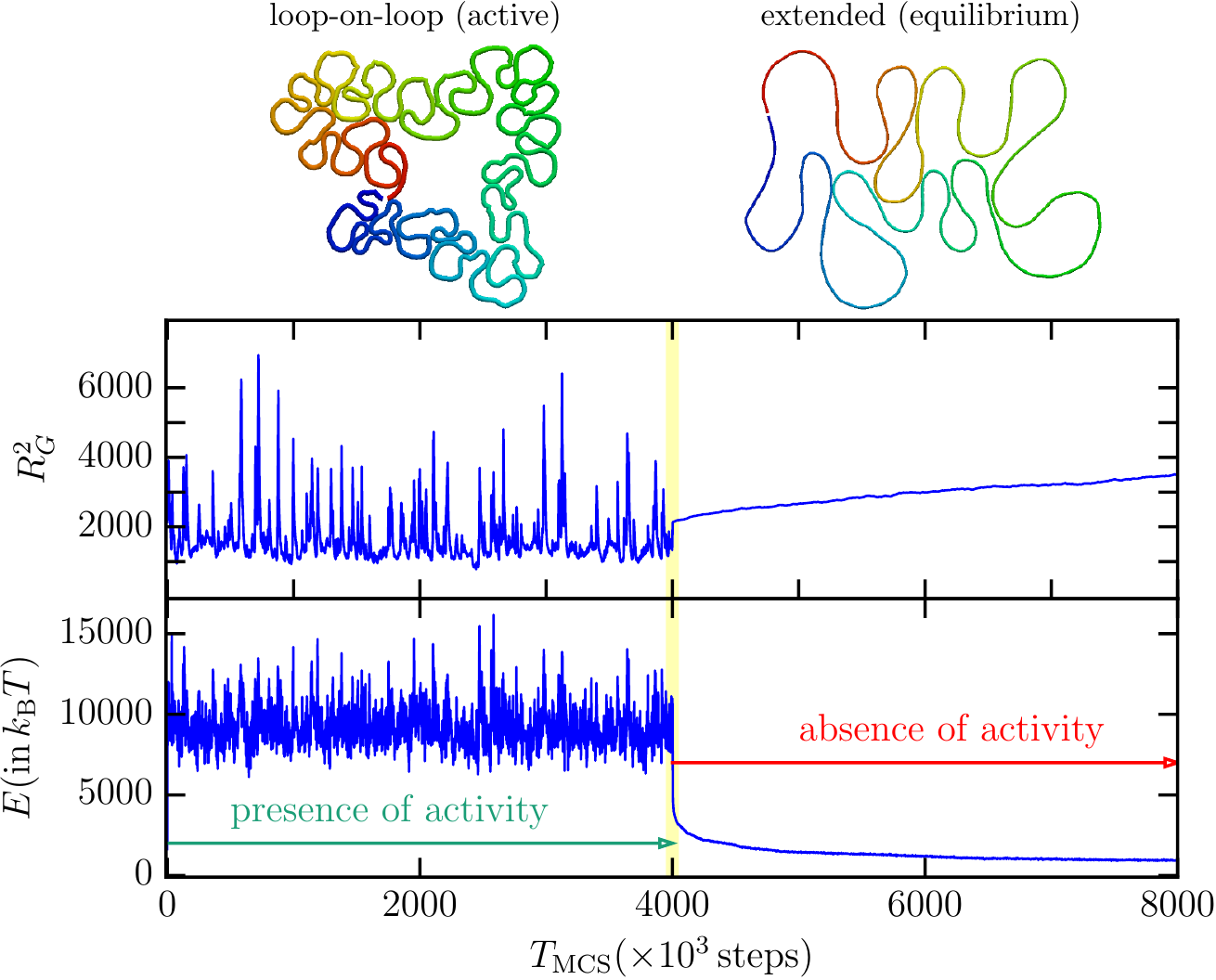}
\caption{\label{fig:polymer-steady-state} The 2d active ring polymer  with $\epsilon_{\pm}=0.1$ \nmcs{} is maintained at its steady state active state for the first half of the simulations ($0-4000\times 10^{3}$ MCS) and the activity is switched off for the later half the simulations ($4000-8000\,\times 10^{3}$ MCS). The top panel shows the representative snapshots of the loop-on-loop and extended conformations of the polymer in the active and equilibrium states respectively. The radius of gyration ($R_{G}^{2}$) and the elastic energy ($E$) show a sudden variation, when the polymer relaxes to its equilibrium state upon switching off activity. Data shown for a polymer with $\kappa=120$ \kbt{}, $\overline{N}=1000$ and $\mu_{\pm}=20$ \kbt{}.}
\end{figure}

The loop-on-loop architecture appear to be characteristic steady state configurations of active  polymers subjected to active addition-removal of beads. Indeed we find loop-on-loop configuration of the active polymer disappears very quickly when activity is suddenly switched off in Fig.~\ref{fig:polymer-steady-state}.

The total simulation time of $8\times10^{6}$ Monte Carlo Steps (MCS) is subdivided into two equal intervals. In the first part of the simulation ($0-4\times10^{6}$MCS) the polymer is subjected to the active bead addition-removal moves (using methods described in Sec.\ref{sec:act-mcsmoves}) while these moves are switched off in the later part ($>4\times10^{6}$MCS) \textemdash~ the corresponding regions in the time axis are shown in Fig.~\ref{fig:polymer-steady-state}. It can be seen that the initial loop-on-loop configuration of the active polymer (top left in Fig.~\ref{fig:polymer-steady-state}) quickly disassembles and gives rise to the equilibrium shape (top right in Fig.~\ref{fig:polymer-steady-state}). 

We also monitor the the total energy of the system ($E$), the radius of gyration ($R_{G}^{2}$) and the fluctuations in the total number of beads ($N$) in the polymer.
These quantities show  sharp changes in behaviour when the activity is switched off (Fig.~\ref{fig:polymer-steady-state}). The transition between the looped to equilibrium shapes and vice-versa continues over any number of cycles, clearly indicating that the compact loop-on-loop shapes is a true steady state of the active polymer.\\

\noindent See {\it Movie-M1.mp4} for a movie of this simulation.
%%%
\section{Statistics of Steady State Configurations of the Active Polymer}

\subsection{Radius of gyration of equilibrium and active polymers}
For a polymer configuration with $N$  beads, the radius of gyration tensor is calculated as,
\begin{equation}
\mathbbm{G}=\frac{1}{N} \sum_{i=1}^{N} \left( \vec{R}_{i}-\vec{R}_{\rm cm} \right) \otimes  \left( \vec{R}_{i}-\vec{R}_{\rm cm} \right).
\end{equation}
$\vec{R}_{i}$ is the position vector of the bead $i$ and $\vec{R}_{\rm cm}=N^{-1}\underset{i} \sum \vec{R}_{i}$ denotes the position of the center of mass of the polymer. Let $\lambda_{1},\,\lambda_{2},$ and $\lambda_{3}$ be the eigenvalues of the gyration tensor $\mathbbm{G}$, with $\lambda_{1}<\lambda_{2}<\lambda_{3}$, then the squared radius of gyration is calculated as,
\begin{equation}
R_{G}^{2}=\overset{d}{\underset{i=1}\sum} \lambda_{i},
\end{equation}
where $d=2$ for a two dimensional polymer and $d=3$ for a three dimensional polymer. The plot of $R_G^2$ for the equilibrium and active polymers (flexible and semi-flexible) in two and three dimensions is given in the main manuscript.

\subsection{Density correlation functions of equilibrium and active polymers}
The real-space number density at a spatial location ${\bf r}$ is defined as  
\begin{equation}
\rho({\bf r})=\sum_{i} \delta\left({\bf r}-{\bf r}_{i}\right),
\end{equation}
where ${\bf r}_{i}$ is the position of the $i^{th}$ bead, and the density-density correlation function between two points separated by a distance $r=|{\bf r}|$ is defined as
\begin{equation}
C_{\rho \rho}(r)= \dfrac{\rho(0)\rho({\bf r})}{N^{2}}.
\end{equation}
\begin{figure}[!h]
\centering
\includegraphics[width=16cm,clip]{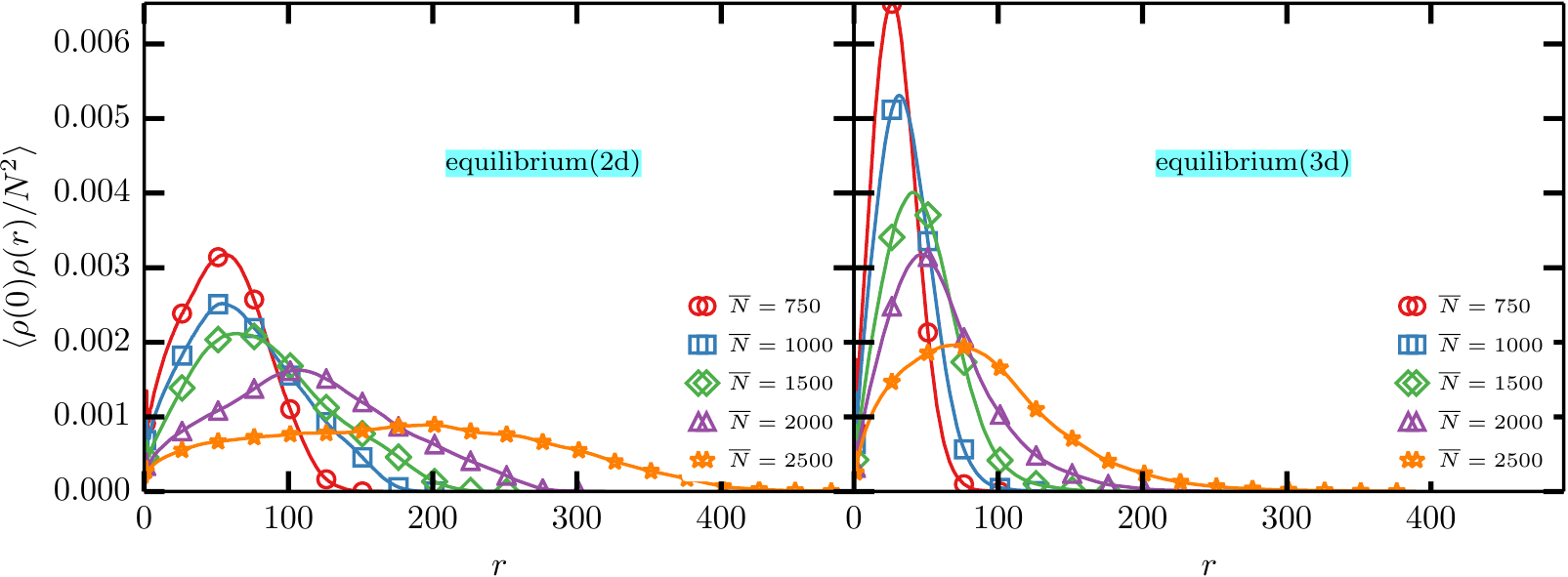}
\includegraphics[width=16cm,clip]{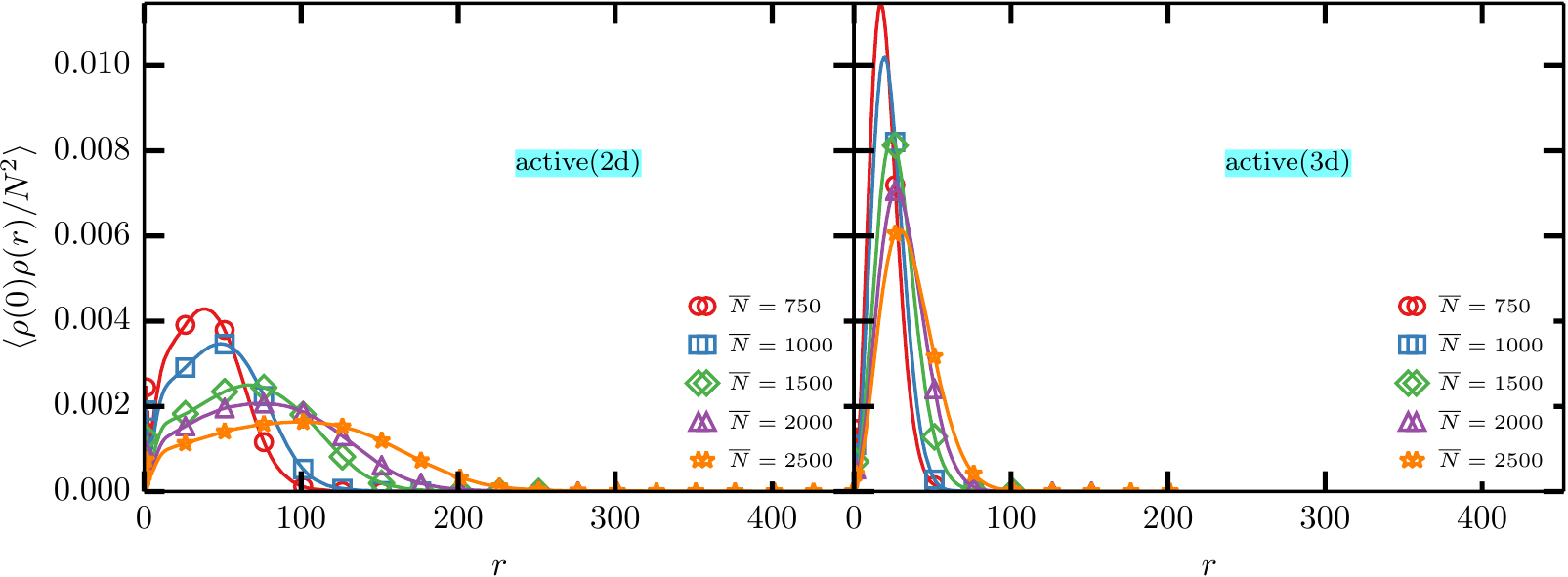}
\caption{\label{fig:rhorhocorr} Density-density correlation of an  equilibrium ring polymer (top panel) and an active ring polymer (bottom panel) in two and three dimensions (left and right), for polymers with $\kappa=120$ \kbt{}, $\epsilon_{\pm}=0.1$ \nmcs, $\mu_{+}=\mu_{-}=20$ \kbt{}, and mean sizes $\overline{N}=750$, $1000$, $1500$, $2000$ and $2500$. Symbols are shown only at a limited number of points.}
\end{figure}

\begin{figure}[!h]
\centering
\includegraphics[width=16cm,clip]{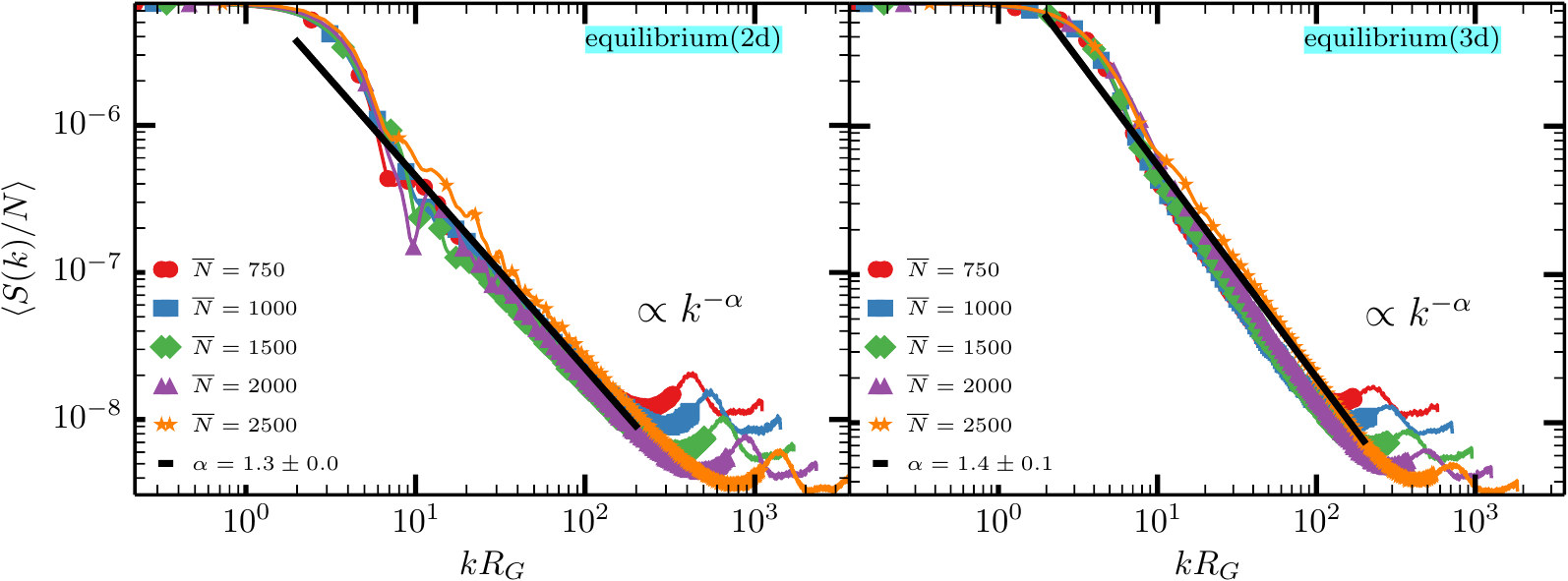}
\includegraphics[width=16cm,clip]{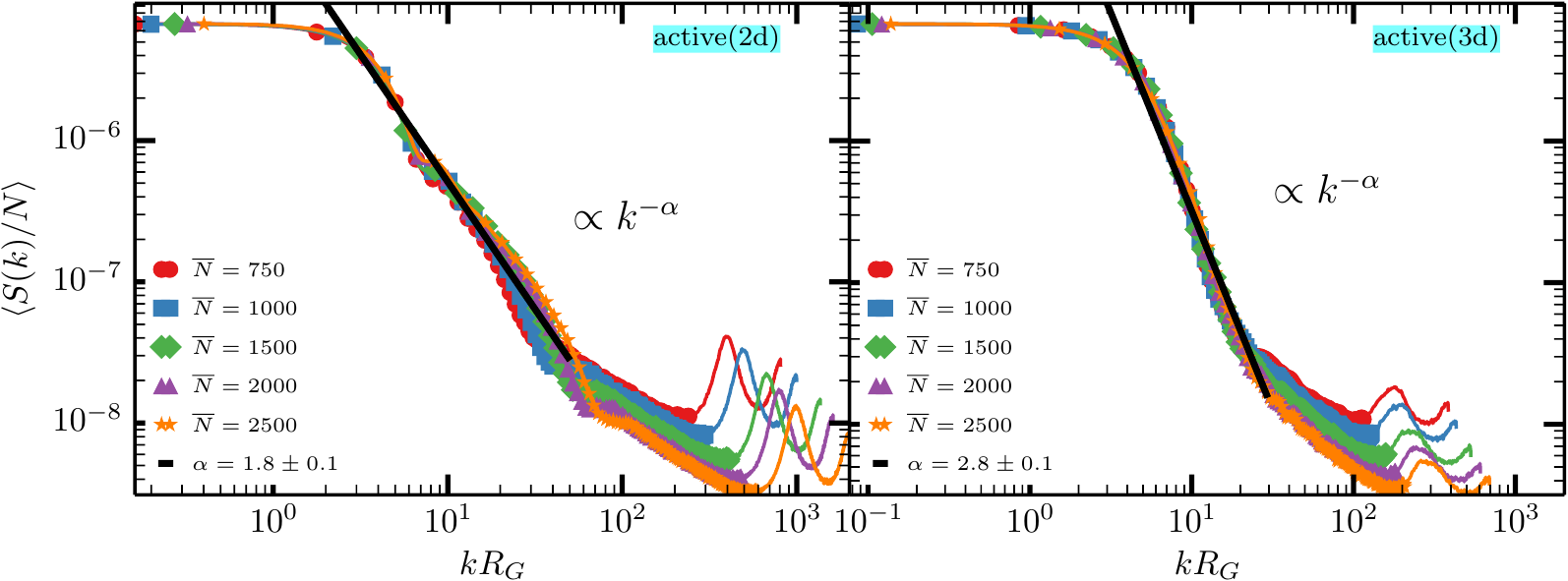}
\caption{\label{fig:struct-fac} The scaled structure factor $S(k)/N$, the Fourier transform of $C_{\rho\rho}(r)$, as a function of $kR_{G}$ for an equilibrium and active ring polymer (top and bottom panels) in two and three dimensions (left and right). Data shown for polymer with  $\kappa=120$ \kbt{}, $\epsilon_{\pm}=0.1$ \nmcs, $\mu_{+}=\mu_{-}=20$ \kbt{}, and mean sizes $\overline{N}=750$, $1000$, $1500$, $2000$ and $2500$. At intermediate values of $kR_{G}$, the structure factor for the active polymer shows a power law behaviour with the power law exponents $\alpha=1.803 \pm 0.106$ in two dimensions and $\alpha=2.803 \pm 0.064$ in three dimensions. The computed values of $\alpha$ agrees with the previously reported values  of $2$ and $3$, for fractal globule polymers~\cite{Halverson:2014hg}, in two and three dimensions, respectively.}
\end{figure}

The density-density correlation for an active polymer in two and three dimensions is shown in Fig. \ref{fig:rhorhocorr}, for polymers with $\overline{N}=750$, $1000$, $1500$, $2000$ and $2500$.  As can be seen from Fig. 3 in the main manuscript the active polymer shows a highly compact organization in both two and three dimensions. 
It can be seen from Fig. \ref{fig:rhorhocorr} that the peak of the density correlation shifts to a larger value of $r$ with increase in the system size. The corresponding plots for the equilibrium polymer are also shown for comparison. \\

The structure factor $S(k)$ for a polymer can be computed as the Fourier transform of the density-density correlation function. The structure factor scaled with system size $N$ is shown in Fig. \ref{fig:struct-fac} as a function of the non-dimensional wavenumber $kR_{G}$, which shows a good data collapse for small and intermediate ranges of $kR_{G}$. The scaled structure factor shows a power law decay for intermediate values of $kR_{G}$, with the best fit power law exponents being $\alpha=1.803 \pm 0.106$ and $2.803 \pm 0.064$ in two and three dimensions. These values  are in agreement with the reported power law exponents for the fractal globule model for chromatin which are $2$ and $3$ in two and three dimensions, respectively~\cite{Halverson:2014hg}.

\subsection{Contour vs geometric length }
\begin{figure}[!h]
\centering
\includegraphics[width=15cm]{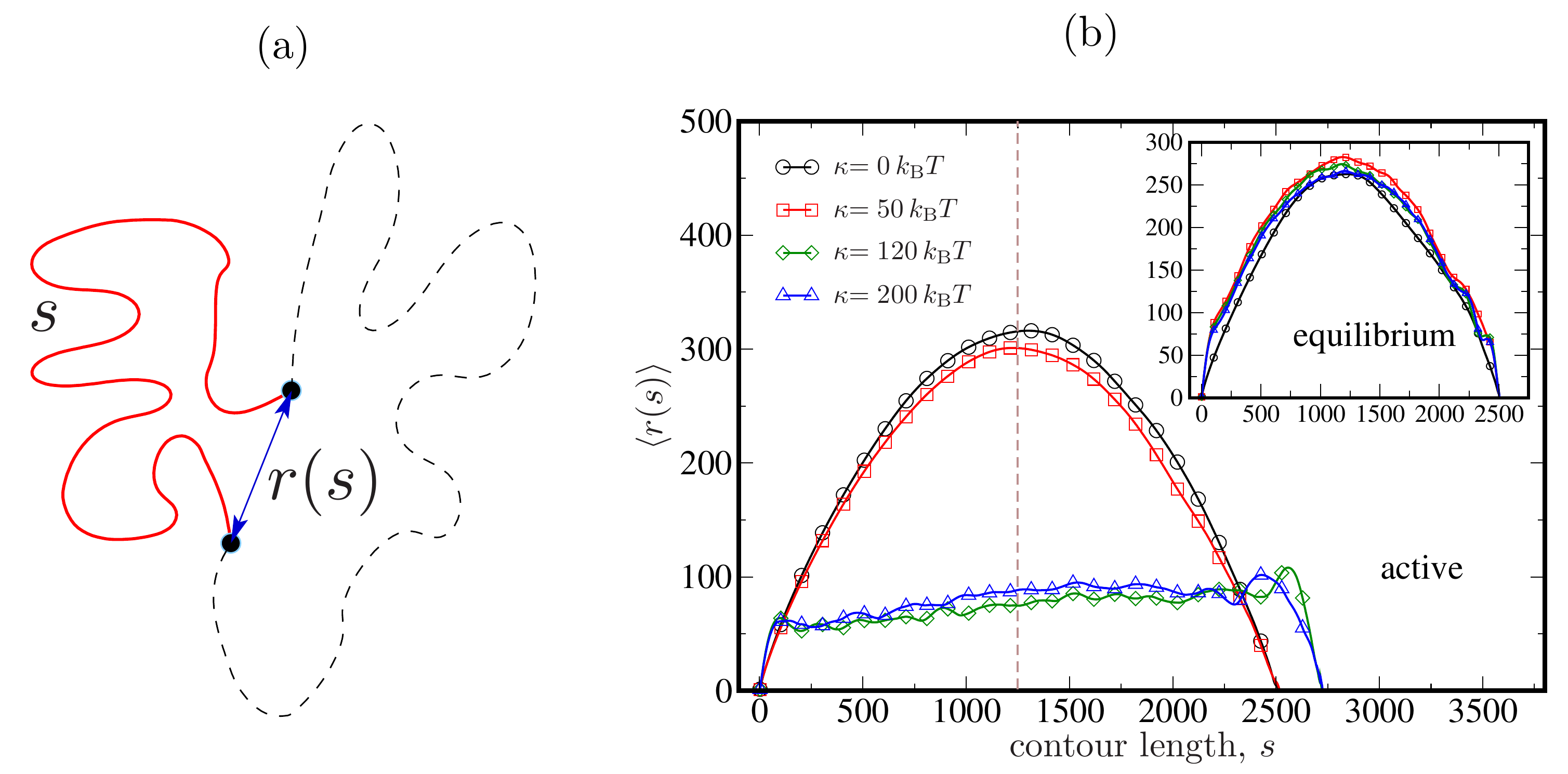}
	
\caption{\label{fig:gegedist}(a) An illustration of the contour length ($s$) and geometric length ($r(s)$) in a polymer segment, and (b) $\langle r(s) \rangle$ vs $s$ as a function of $\kappa$ for an equilibrium (inset) and an active ring polymer in 2d (main panel), with $N=2500$, $\epsilon_{\pm}=0.1$ \nmcs, and $\mu_{+}=\mu_{-}=20$ \kbt{}. The vertical line  marks the position of half the polymer length.}
\end{figure}

The contour length $s$ of a polymer segment connecting two beads $i$ and $j$ is proportional to $|i-j|$, however the corresponding geometric length $r(s)$ depends on the entire polymer conformation in between the beads. Contour and geometric lengths, for an arbitrary polymer segment, is shown in Fig. \ref{fig:gegedist}(a). In the case of an extended polymer it can easily be shown that $r(s) \approx s$, for $s<L/2$ where $L$ is the total length of the  segment under study. 

The average geometric distance $\langle r(s) \rangle$ as a function of the segment length $s$, for both active and equilibrium 2d polymers, for various bending stiffness is shown in Fig. \ref{fig:gegedist}(b). The geometric length of the equilibrium polymer, shown in the inset to Fig.~\ref{fig:gegedist}(b),  for three values of $\kappa$, increases with $s$ and  is peaked at $s=N/2$. As expected for  self avoiding flexible ($\kappa =0$) polymers, $\langle r(s) \rangle \sim s^{3/4}$, while stiff polymers with non-zero value of  $\kappa$ show a nearly linear scaling.  An active polymer in the semi flexible regime ($\kappa=120$ \kbt{}) however shows a different trend \textemdash~ $\langle r(s) \rangle$ stays relatively flat as a function of $s$  as is seen in Fig. \ref{fig:gegedist}(b). The flatness in the distribution of the geometric length  suggests that segments far separated along the polymer backbone 
come into spatial proximity of each other, and that this contact events are very dynamic.
This proximal nature can be best understood from the contact  probability distribution $P_{\rm con}(s)$.

\subsection{Occurrence probability and contact map of equilibrium and active polymers}
For a given polymer segment of length $s$, the probability for occurrence of all values of $r(s)$ can be useful in understanding the compactness of polymers - this measure will be called the occurrence probability of $r(s)$, represented by $P(r(s))$.  It is the probability to find the ends of a polymer segment with contour length $s$ at a geometric separation $r(s)$.  The occurrence probability map or contact map for a given polymer conformation is defined as,
\begin{equation}
{\cal P}(r(s))=\dfrac{1}{2} \overset{N}{\underset{i=1} \sum}\overset{N}{\underset{j=1} \sum}\, \delta_{s,s_{ij}}\, \delta_{r(s),r_{ij}},
\end{equation}
where $r_{ij}=\vert {\bf R}_i-{\bf R}_j|$, ${\bf R}_{i}$ is the position of bead $i$  and $s_{ij}$ is the contour length  of the polymer between beads $i$ and $j$, and $\delta_{\alpha\beta}$ is the Kronecker delta function. The contact probability density is defined as
\begin{equation}
P(r(s))=\dfrac{{\cal P}(r(s))}{\underset{s} \sum \underset{r(s)} \sum {\cal P}(r(s))}.
\end{equation}

\begin{figure}[!h]
\centering
\includegraphics[width=15cm,clip]{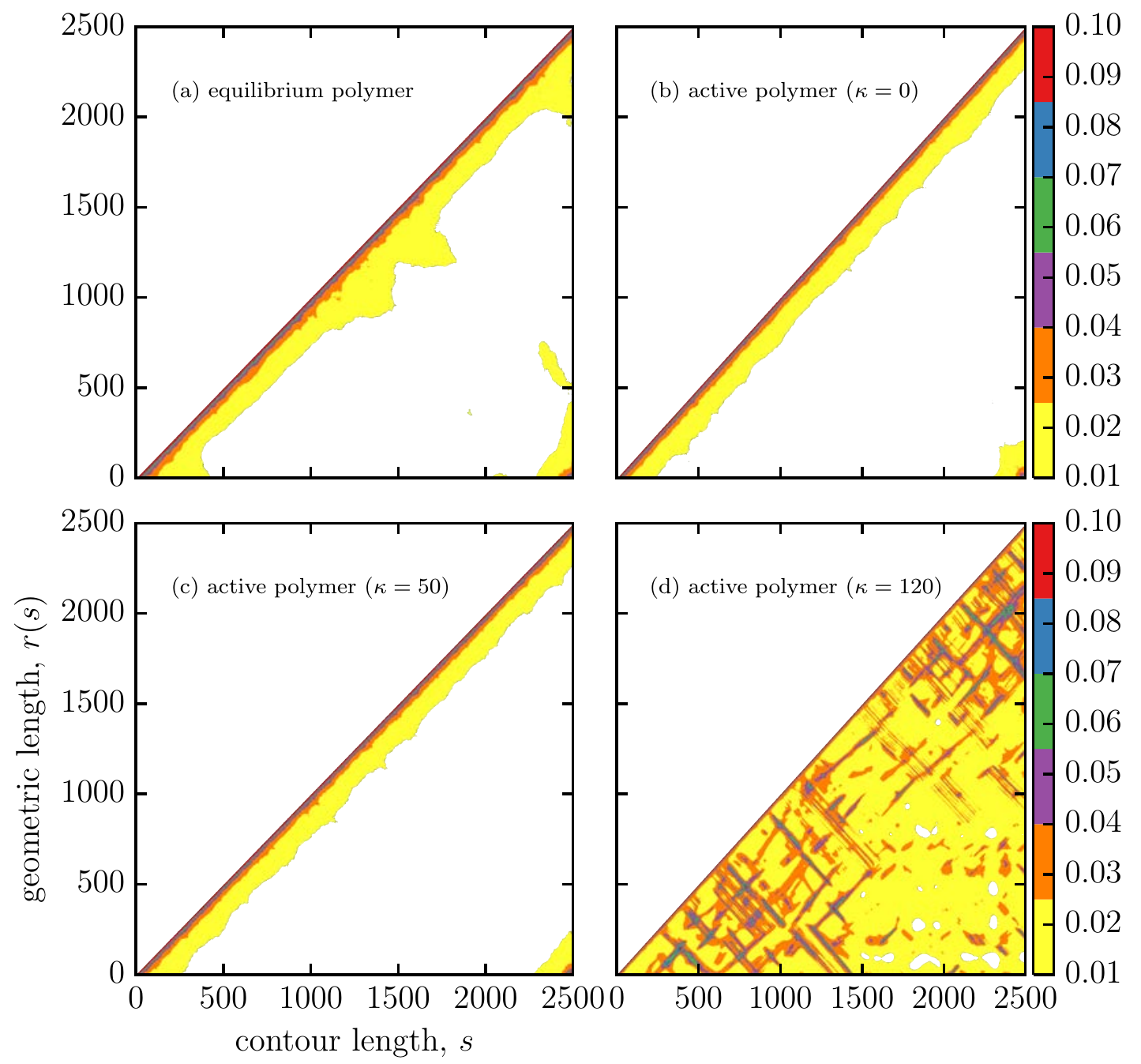}
\caption{\label{fig:contactmap} Contact maps for the conformations of (a) an equilibrium 2d ring polymer with $\kappa=0$, (b, c, d) active 2d ring polymers with $\kappa=0$, $50$ \kbt{}, and $120$ \kbt{}, at fixed values of $\overline{N}=2500$, $\epsilon_{\pm}=0.1$ \nmcs, and $\mu_{+}=\mu_{-}=20$\kbt{}. Extended polymer configurations, characteristic of panel (a), (b) and (c) show short range contacts which appear here as a narrow band in the contact map.  On the other hand, active polymers with loop-on-loop configurations, which is characteristic of panel (d), show long range contacts which appears as a diffuse profile in the contact map. } 
\end{figure}

The contact maps for an equilibrium polymer with $\kappa=0$ is given in Fig. \ref{fig:contactmap}(a) and that for active polymers with $\kappa=0$, $50$, and $120$ \kbt{} are shown in  Figs.~\ref{fig:contactmap}(b-d).

The occurrence probability map shows a marked difference in the contact profile of equilibrium and active polymer. Equilibrium polymer configurations have relatively short range contacts which shows up as a thin diagonal band in the contact probability map, the range of these interactions depend on the persistence length of the equilibrium polymer. Thus contact map for 2d equilibrium polymers with $\kappa=0$, has a slightly more diffuse appearance compared to the configurations of equilibrium polymer with $\kappa >0$,
which appear as a thin diagonal band. In contrast the contact map of the 2d active polymer
with $\kappa=0$ and $50$ \kbt{} shows extended configurations and a thin band (see panels (a), (b), and (c) in Fig.~\ref{fig:contactmap}), whereas the contact map of the active polymer with $\kappa=120$ \kbt{}, given in Fig. \ref{fig:contactmap}(d), shows a very broad distribution that reveals the existence of non zero contacts for all values of $s$ and $r(s)$.

\begin{figure}
\centering
\includegraphics[width=15cm,clip]{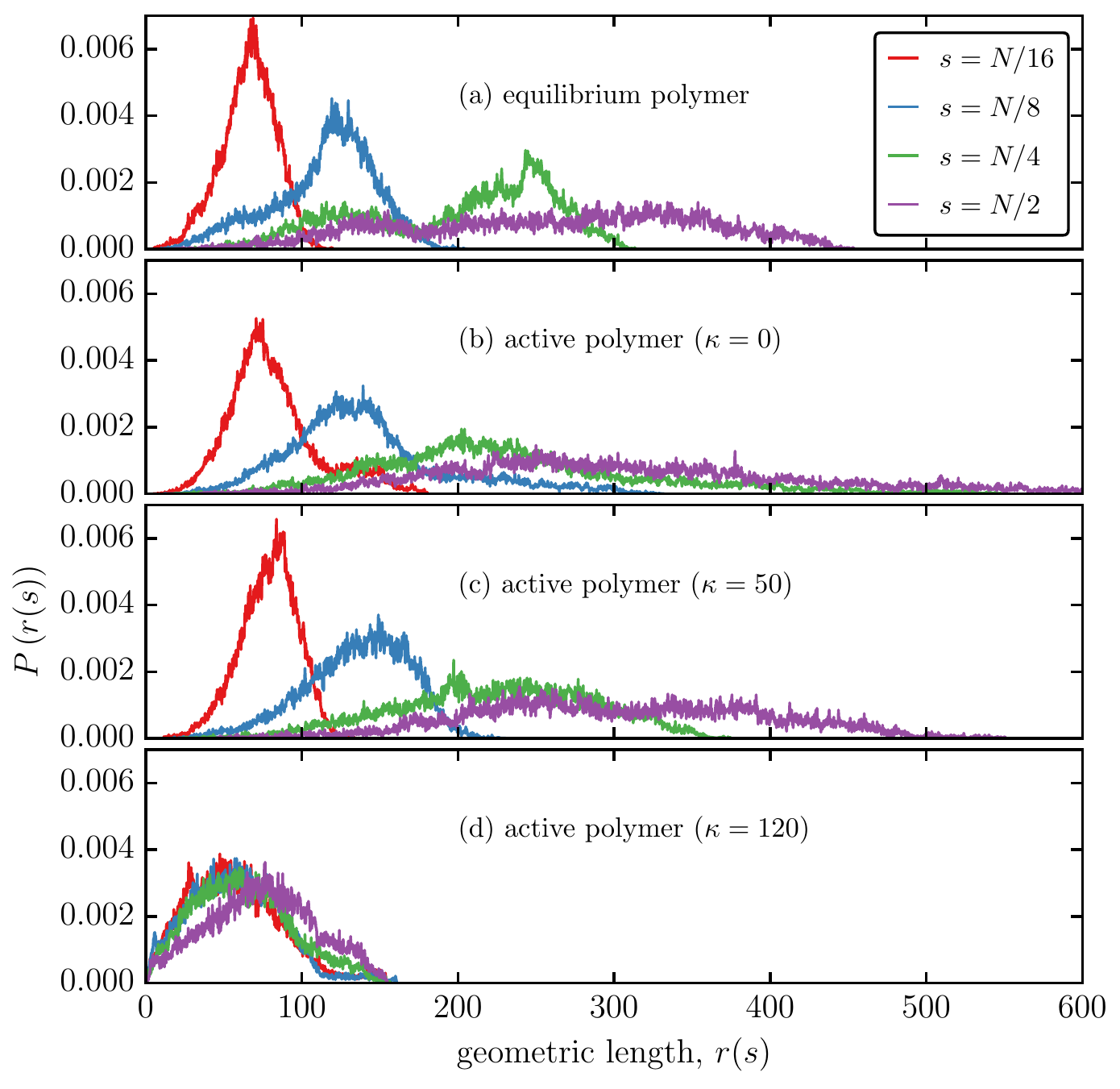}
\caption{\label{fig:geomdist-funcN} Distribution of the geometric distance $r(s)$, for four values of contour length $s=N/16,\,N/8,\,N/4,\,N/2$  with $N=2500$. Data shown in panels (a), (b), (c) and (d) corresponds to heat map data shown in Figs. \ref{fig:contactmap}(a), (b), (c), and (d) respectively.}
\end{figure}

%\begin{enumerate}[(i)]
%\item Extended polymer configurations have short range contacts, which appears as a thin diagonal band in panels (a), (b), and (c) in Fig.~\ref{fig:contactmap}. However, the equilibrium polymer with $\kappa=0$, has longer range contacts compared to that seen for stiff equilibrium polymers (data not shown) and for active polymers with extended configurations.

The compactness of the 2d active polymer is clearly seen when  one looks at the occurrence probability as a function of geometric distance $r(s)$ at a fixed length of the polymer segment. $P(r(s))$ corresponding to segment lengths $s=N/16$, $N/8$, $N/4$ and $N/2$ is shown in Fig. \ref{fig:geomdist-funcN} for an equilibrium polymer (panel (a))  and for an  active polymer with $\kappa=0$, $50$, and $120$ \kbt{} (panels (b)-(d)).

For all values of $s$ analyzed here, $P(r(s))$ for an equilibrium flexible ($\kappa=0$) polymer shows a symmetric unimodal distribution, see Fig. \ref{fig:geomdist-funcN}(a),  whose width increases with increasing $s$ \textemdash~ the peak value of $P(r(s))$ represents the characteristic length scale that dominates the spatial conformation of the polymer. $P(r(s))$ becomes uniformly distributed with increase in $s$ and this behavior persists even for active polymers with $\kappa=0$ and $\kappa=50$ \kbt{}, see Figs. \ref{fig:geomdist-funcN}(b, c). However there is a significant deviation in the profile of $P(r(s))$ when the active polymer forms loop-on-loop configurations, as shown for the case of $\kappa=120$ \kbt{} in panel (d). The unimodal distribution persists for all values of $s$ considered here and also peaks at the same value of $r(s)$, which implies that the polymer is compact at all scales and activity sets a characteristic length scale $\xi_{a}$, as discussed in the main manuscript.

\subsection{Contact probability of equilibrium and active polymers}

Two beads separated by a distance  $s$ along the polymer contour, are said to be in contact if their geometric distance $r(s)$ is within one tether length ( $a_{0}<r(s)<\sqrt{3}a_{0}$), which is the self avoidance constraint defined in main text and in Sec.~\ref{sec:act-mcsmoves}. \\

The contact probability  $P_{\rm con}(s)$ is computed from the occurrence probability as,
\begin{equation}
P_{\rm con}(s)=\underset{s}\sum \underset{r(s)}\sum P(r(s)) {\bf\Theta}(r(s)-a_{0})(1-{\bf \Theta}(r(s)-\sqrt{3}a_{0})),
\end{equation}
where ${\bf \Theta}(r(s)-a_{0})$ is the unit Heaviside step function, which is $0$ when $r(s)<a_{0}$ and $1$  when $r(s)>a_{0}$.

The contact probability of two interior points of a self avoiding, open, flexible, polymer chain in two dimensions has been shown to obey the scaling,
\begin{equation}
P_{\rm con}(s) \sim s^{-\theta},
\end{equation}
where the scaling exponent  $\theta =43/16 \sim 2.6785$~\cite{Duplantier:1987vw}. We have verified this scaling for the equilibrium self-avoiding polymer at $\kappa=0$
(Fig. 4(b) of main manuscript). The scaling behavior of the contact probability for active polymers has also been discussed in the main manuscript.

\subsection{Calculation of the loop size distribution $P_{\rm loop}(s)$}
We can also compute the distribution of loop sizes, measured along the contour length $s$. Two beads $i$ and $j$ separated by a contour length $s$ along the polymer form a  loop if their
\begin{enumerate}[(a)]
\item  geometric distance $R_{ij}=|\vec{R}_i-\vec{R}_j|$ is such that $a_{0}<R_{ij}<\sqrt{3}a_{0}$, and
\item  their tangent vectors subtend an angle $\theta$ with each other, such that   $\theta=\cos^{-1} \left(\hat{t}_{i} \cdot \hat{t}_{j} \right) > 3\pi/4$.
\end{enumerate}

\begin{figure}[!h]
\centering 
\includegraphics[width=4in]{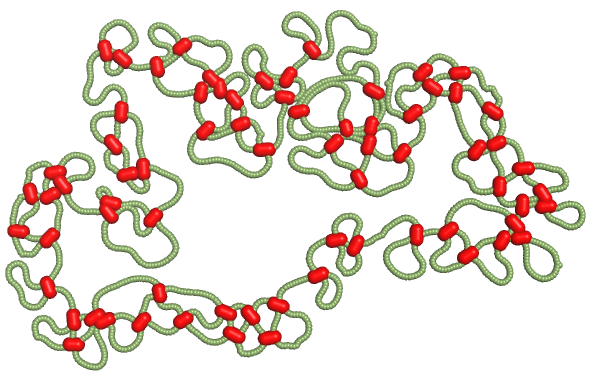}
\caption{\label{fig:loopexample} An active 2d ring polymer with all possible contact points on it marked explicitly. The red colored cross linkers connects two polymer beads forming a loop.}
\end{figure}

Based on the above rules, we compute loops of all possible lengths at steady state. Fig. \,\ref{fig:loopexample} show a two dimensional active polymer, with $N=2500$, with all possible contact points explicitly marked. The loop size distribution for active polymers in 3d has a very broad distribution, suggesting that there is no mean loop
size (Fig.4(a) of main manuscript).\\

\subsection{Segmental Activity : Segregation into territories}
We study the shape of an active polymer when subjected to spatially inhomogeneous activity, as might be expected in an in-vivo context. This is motivated by the fact that the euchromatin and heterochromatin regions do not show the same folded configurations. For this we compartmentalize a linear polymer, with $\overline{N}=2500$ into three segments. The region defining the middle segment is kept inactive for the entire duration of the simulation and the other two segments are subjected to activity with $\epsilon_{\pm}=0.1$ \nmcs, $\mu_{+}=\mu_{-}=20$ \kbt{}, and $\kappa=120$ \kbt{}. The number of active segments in the polymer is denoted by ${\cal N}_{\rm seg}$.

\begin{figure}[!h]
	\centering
	\includegraphics[width=3in]{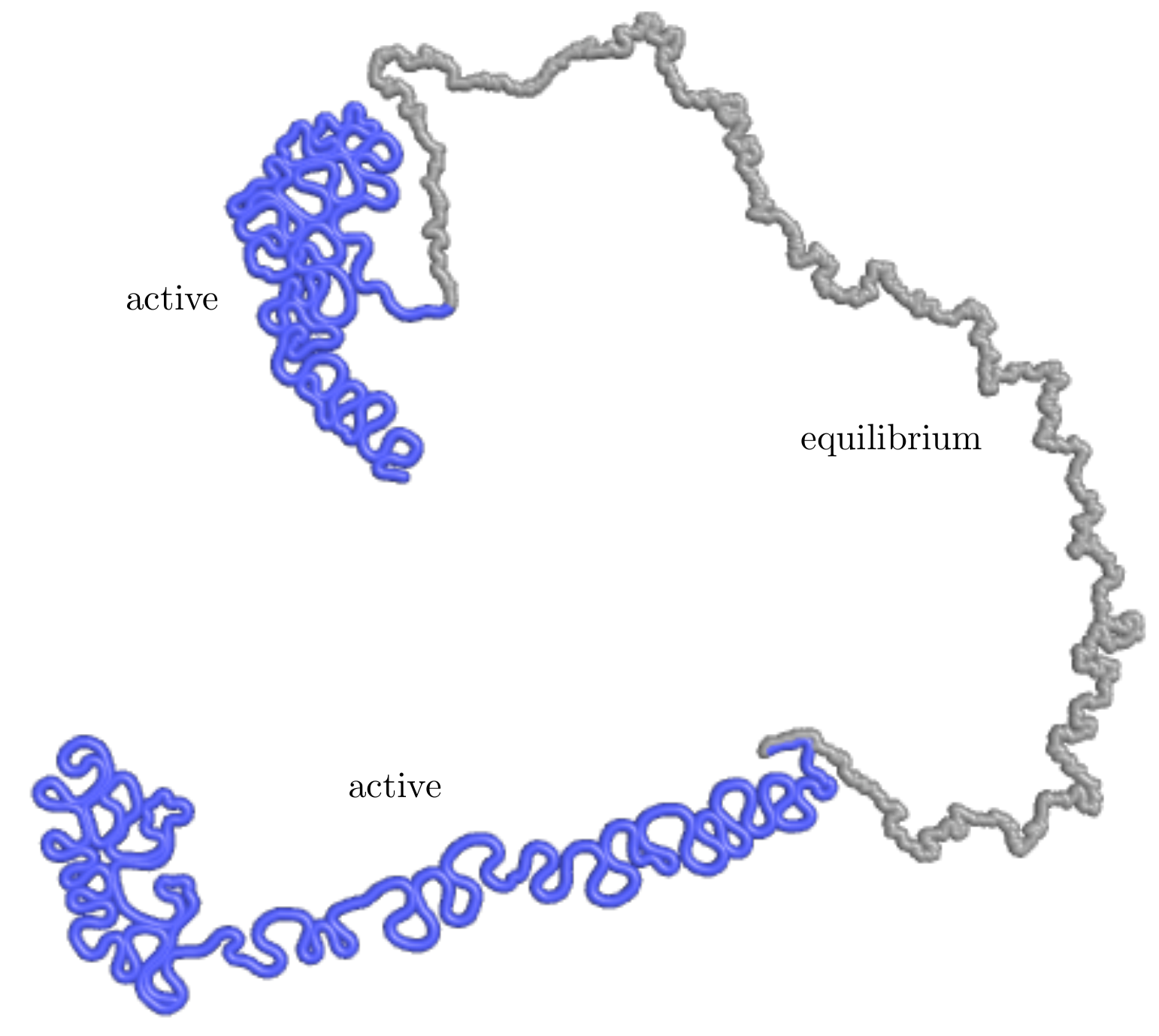}
	\caption{\label{fig:3seg} An inactive semiflexible polymer segment sandwiched between two active segments in a linear 2d polymer.}
\end{figure}

Fig.~\ref{fig:3seg} shows the resulting steady state shape for a two dimensional polymer with ${\cal N}_{\rm seg}=2$ . The regions subjected to activity form folded configurations, similar to those seen in a fully active polymer, and remain compact. The inactive region remains in an extended conformation as expected of a semi-flexible polymer in equilibrium.

\begin{figure}[!h]
\centering
\includegraphics[width=6in]{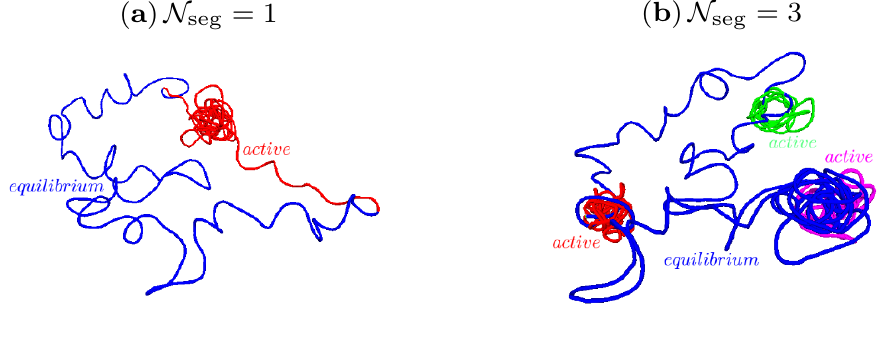}
\caption{\label{fig:3dseg} Snapshots of a three dimensional ring polymer with {\bf (a)} ${\cal N}_{\rm seg}=1$  and {\bf (b)} ${\cal N}_{\rm seg}=3$  active segments. The active moves have been performed with $\mu_{+}=\mu_{-}=20$ \kbt{}, and $\epsilon_{\pm}=0.1$ \nmcs, and $\kappa=120$ \kbt{}. 	}
\end{figure}

We perform similar studies in three dimensions using a ring polymer and the snapshots of the polymer with ${\cal N}_{\rm seg}=1$ and ${\cal N}_{\rm seg}=3$ active segments are shown in Fig.~\ref{fig:3dseg}(a) and (b) respectively.
% We also compare the contact probabilities($P_{\rm con}(s)$) of the segmented active polymers to a fully active polymer in Fig.~\ref{fig:3dseg}(c). It can be seen that the contact probability increases with increase in the number of active beads and the contact probability for a fully active polymer scales as $s^{-3/2}$ and tends to the equilibrium limit of $s^{-1}$ with increase in the number of equilibrium beads. 
The corresponding movies are shown in movies M2 and M3.

%The contact probability map for the fully active polymer and the segmented active polymer is shown in Fig.~\ref{fig:3dseg-heatmap}. While the fully active polymer displays long range contact for its entire contour length, the contact profile for the segmented polymer is dependent on the number of active segments. It can be seen that the long range contact profile has one, two, three bright bands corresponding to ${\cal N}_{\rm seg}=1,\,2,$ and $3$.

%\begin{figure}[!h]
%\centering
%\includegraphics[width=6in]{./figure-s21}
%\caption{\label{fig:3dseg-heatmap} The contact probability density $P(s,r(s))$ for a three dimensional polymer with fully active, and partly active segments -- data shown for ${\cal N}_{\rm seg}=1,\,2,$ and $3$ active segments. The active moves have been performed with $\mu_{+}=\mu_{-}=20$ \kbt{}, and $\epsilon_{\pm}=0.1$ \nmcs.}
%\end{figure}

\subsection{Movie M1: Steady state nature of the loop on loop shapes}
The movie illustrates the steady state nature of the loop on loop shapes of the active polymer with $\overline{N}=1000$,  $\kappa=120$ \kbt{}, $\Delta p=0$, $\mu_{+}=\mu_{-}=20$ \kbt{} and $\epsilon_{+}=\epsilon_{-}=0.1$  \nmcs{}. The simulations are run for a total of 8 million MCS, of which the active moves are performed only for the first 4 million MCS. It can be seen that the compact shape of the polymer is characteristic of non-equilibrium length fluctuations in the polymer. The highly curved shape of the active polymer becomes unstable and the polymer relaxes back to its equilibrium shapes for all times $T_{MCS}>4\times 10^{6}$ MCS.

\subsection{Movies M2 and M3: Evolution of a 3d ring polymer with coexisting active and equilibrium segments}
A ring polymer is divided into 2${\cal N}_{\rm seg}$  equal sized segments and ${\cal N}_{\rm seg}$ segments of these are marked to be active while the rest are treated to be at equilibrium. In the presence of activity, the active segments shows a collapsed conformation while the equilibrium segments show an extended conformation, consistent with our earlier findings. Movies M2 and M3 correspond to a polymer with ${\cal N}_{\rm seg}=1$ and ${\cal N}_{\rm seg}=3$. The parameters for the active segments are set to be $\kappa=120$\kbt{}, $\mu_{+}=\mu_{-}=20$ \kbt{}, and $\epsilon_{\pm}=0.1$ \nmcs{}.

%\bibliographystyle{aipnum4-1}
%\bibliography{../bibfile}

\begin{thebibliography}{10}

\bibitem{LiebermanAiden:2009jz}
Lieberman-Aiden E, {et~al.}
\newblock (2009) {Comprehensive mapping of long-range interactions reveals
  folding principles of the human genome.}
\newblock \emph{Science} 326:289--293.

\bibitem{Zhang:2012iu}
Zhang Y, {et~al.}
\newblock (2012) {Spatial organization of the mouse genome and its role in
  recurrent chromosomal translocations.}
\newblock \emph{Cell} 148:908--921.

\bibitem{Sexton:2012cn}
Sexton T, {et~al.}
\newblock (2012) {Three-dimensional folding and functional organization
  principles of the Drosophila genome.}
\newblock \emph{Cell} 148:458--472.

\bibitem{Grosberg:1993fj}
Grosberg A, Rabin Y, Havlin S, Neer A
\newblock (1993) {Crumpled globule model of the three-dimensional structure of
  DNA}.
\newblock \emph{Euro. Phys. Lett.} 23:373.

\bibitem{Mirny:2011cl}
Mirny LA
\newblock (2011) {The fractal globule as a model of chromatin architecture in
  the cell}.
\newblock \emph{Chromosome Res.} 19:37--51.

\bibitem{wolynes:2015}
Zhang B, Wolynes PG
\newblock (2015) {Topology, structures, and energy landscapes of human chromosomes}
\newblock \emph{Proc. Natl. Acad. Sci. U.S.A.} 109:16173--16178.




\bibitem{vandenEngh:1992bi}
van~den Engh G, Sachs R, Trask B
\newblock (1992) {Estimating genomic distance from DNA sequence location in
  cell nuclei by a random walk model}.
\newblock \emph{Science} 257:1410--1412.

\bibitem{Halverson:2014hg}
Halverson JD, Smrek J, Kremer K, Grosberg AY
\newblock (2014) {From a melt of rings to chromosome territories: the role of
  topological constraints in genome folding}.
\newblock \emph{Rep. Prog. Phys.} 77:022601.

\bibitem{Nicodemi:2009hb}
Nicodemi M, Prisco A
\newblock (2009) {Thermodynamic Pathways to Genome Spatial Organization in the
  Cell Nucleus}.
\newblock \emph{Biophys. J.} 96:2168--2177.

\bibitem{Barbieri:2012iw}
Barbieri M, {et~al.}
\newblock (2012) {Complexity of chromatin folding is captured by the strings
  and binders switch model.}
\newblock \emph{Proc. Natl. Acad. Sci. U.S.A.} 109:16173--16178.

\bibitem{Brackley:2013dt}
Brackley CA, Taylor S, Papantonis A, Cook PR, Marenduzzo D
\newblock (2013) \emph{{Nonspecific bridging-induced attraction drives
  clustering of DNA-binding proteins and genome organization}}
\newblock (Scottish Universities Physics Alliance (SUPA), School of Physics,
  University of Edinburgh, Edinburgh EH9 3JZ, United Kingdom.), pp 3605--E3611.

\bibitem{Rosa:2008bp}
Rosa A, Everaers R
\newblock (2008) {Structure and Dynamics of Interphase Chromosomes}.
\newblock \emph{PLoS Comput Biol} 4:e1000153.

\bibitem{Saha:2006do}
Saha A, Wittmeyer J, Cairns BR
\newblock (2006) {Chromatin remodelling: the industrial revolution of DNA
  around histones.}
\newblock \emph{Nat. Rev. Mol. Cell Biol.} 7:437--447.

\bibitem{BrowerToland:2002fr}
Brower-Toland BD, {et~al.}
\newblock (2002) {Mechanical disruption of individual nucleosomes reveals a
  reversible multistage release of DNA.}
\newblock \emph{Proc. Natl. Acad. Sci. U.S.A.} 99:1960--1965.

\bibitem{doi:1986bk}
Doi M, Edwards SF
\newblock (1986) \emph{{The Theory of Polymer Dynamics}}
\newblock (Oxford University Press, USA).

\bibitem{Duan:2010jf}
Duan Z, {et~al.}
\newblock (2010) {A three-dimensional model of the yeast genome}.
\newblock \emph{Nature} 465:363--367.

\bibitem{Hameed:2012ej}
Hameed FM, Rao M, Shivashankar GV
\newblock (2012) {Dynamics of passive and active particles in the cell
  nucleus.}
\newblock \emph{PLoS ONE} 7:e45843.

\bibitem{Bruinsma:2014jv}
Bruinsma R, Grosberg AY, Rabin Y, Zidovska A
\newblock (2014) {Chromatin Hydrodynamics}.
\newblock \emph{Biophys. J.} 106:1871--1881.

\bibitem{Weber:2012gd}
Weber SC, Spakowitz AJ, Theriot JA
\newblock (2012) {Nonthermal ATP-dependent fluctuations contribute to the in
  vivo motion of chromosomal loci.}
\newblock \emph{Proc. Natl. Acad. Sci. U.S.A.} 109:7338--7343.

\bibitem{Ghosh:2014ky}
Ghosh A, Gov NS
\newblock (2014) {Dynamics of Active Semiflexible Polymers}.
\newblock \emph{Biophys. J.} 107:1065--1073.

\bibitem{Talwar:2013bh}
Talwar S, Kumar A, Rao M, Menon GI, Shivashankar GV
\newblock (2013) {Correlated spatio-temporal fluctuations in chromatin
  compaction states characterize stem cells.}
\newblock \emph{Biophys. J.} 104:553--564.

\bibitem{Ganai:2014kv}
Ganai N, Sengupta S, Menon GI
\newblock (2014) {Chromosome positioning from activity-based segregation}.
\newblock \emph{Nucleic Acids Research}.

\bibitem{Kratky:1949un}
Kratky O, Porod G
\newblock (1949) {Diffuse small-angle scattering of X-rays in colloid systems.}
\newblock \emph{J Colloid Sci} 4:35--70.

\bibitem{Leibler:1987eq}
Leibler S, Singh R, Fisher M
\newblock (1987) {Thermodynamic behavior of two-dimensional vesicles.}
\newblock \emph{Phys. Rev. Lett.} 59:1989--1992.

\bibitem{dekker:2013hi}
Dekker J, Marti-Renom MA, Mirny LA
\newblock (2013) {Exploring the three-dimensional organization of genomes:
  interpreting chromatin interaction data}.
\newblock \emph{Nat Rev Genet} 14:390--403.

\bibitem{Frenkel:2001}
Frenkel D, Smit B
\newblock (2001) \emph{{Understanding Molecular Simulation : From Algorithms to
  Applications}}
\newblock (Academic Press), 2 edition.

\bibitem{MateosLangerak:2009go}
Mateos-Langerak J, {et~al.}
\newblock (2009) {Spatially confined folding of chromatin in the interphase
  nucleus.}
\newblock \emph{Proc. Natl. Acad. Sci. U.S.A.} 106:3812--3817.

\bibitem{Duplantier:1987vw}
Duplantier B
\newblock (1987) {Exact contact critical exponents of a self-avoiding polymer
  chain in two dimensions.}
\newblock \emph{Phys. Rev. B} 35:5290--5293.

\bibitem{Virnau:2006jt}
Virnau P, Mirny LA, Kardar M
\newblock (2006) {Intricate Knots in Proteins: Function and Evolution}.
\newblock \emph{PLoS Comput Biol} 2:e122.

\bibitem{Faloutsos:1989wn}
Faloutsos C, Roseman S
\newblock (1989) \emph{{Fractals for secondary key retrieval}}.

\bibitem{Jagadish:1990bu}
Jagadish HV
\newblock (1990) \emph{{Linear Clustering of Objects with Multiple Attributes}}
\newblock (ACM, New York, NY, USA), pp 332--342.

\end{thebibliography}

\begin{thebibliography}{4}%
	\makeatletter
	\providecommand \@ifxundefined [1]{%
		\@ifx{#1\undefined}
	}%
	\providecommand \@ifnum [1]{%
		\ifnum #1\expandafter \@firstoftwo
		\else \expandafter \@secondoftwo
		\fi
	}%
	\providecommand \@ifx [1]{%
		\ifx #1\expandafter \@firstoftwo
		\else \expandafter \@secondoftwo
		\fi
	}%
	\providecommand \natexlab [1]{#1}%
	\providecommand \enquote  [1]{``#1''}%
	\providecommand \bibnamefont  [1]{#1}%
	\providecommand \bibfnamefont [1]{#1}%
	\providecommand \citenamefont [1]{#1}%
	\providecommand \href@noop [0]{\@secondoftwo}%
	\providecommand \href [0]{\begingroup \@sanitize@url \@href}%
	\providecommand \@href[1]{\@@startlink{#1}\@@href}%
	\providecommand \@@href[1]{\endgroup#1\@@endlink}%
	\providecommand \@sanitize@url [0]{\catcode `\\12\catcode `\$12\catcode
		`\&12\catcode `\#12\catcode `\^12\catcode `\_12\catcode `\%12\relax}%
	\providecommand \@@startlink[1]{}%
	\providecommand \@@endlink[0]{}%
	\providecommand \url  [0]{\begingroup\@sanitize@url \@url }%
	\providecommand \@url [1]{\endgroup\@href {#1}{\urlprefix }}%
	\providecommand \urlprefix  [0]{URL }%
	\providecommand \Eprint [0]{\href }%
	\providecommand \doibase [0]{http://dx.doi.org/}%
	\providecommand \selectlanguage [0]{\@gobble}%
	\providecommand \bibinfo  [0]{\@secondoftwo}%
	\providecommand \bibfield  [0]{\@secondoftwo}%
	\providecommand \translation [1]{[#1]}%
	\providecommand \BibitemOpen [0]{}%
	\providecommand \bibitemStop [0]{}%
	\providecommand \bibitemNoStop [0]{.\EOS\space}%
	\providecommand \EOS [0]{\spacefactor3000\relax}%
	\providecommand \BibitemShut  [1]{\csname bibitem#1\endcsname}%
	\let\auto@bib@innerbib\@empty
	%</preamble>
	\bibitem [{\citenamefont {Frenkel}\ and\ \citenamefont
		{Smit}(2001)}]{Frenkel:2001}%
	\BibitemOpen
	\bibfield  {author} {\bibinfo {author} {\bibfnamefont {D.}~\bibnamefont
			{Frenkel}}\ and\ \bibinfo {author} {\bibfnamefont {B.}~\bibnamefont {Smit}},\
	}\href {http://www.worldcat.org/isbn/0122673514} {\emph {\bibinfo {title}
		{{Understanding Molecular Simulation : From Algorithms to Applications}}}},\
\bibinfo {edition} {2nd}\ ed.\ (\bibinfo  {publisher} {Academic Press},\
\bibinfo {year} {2001})\BibitemShut {NoStop}%
\bibitem [{\citenamefont {Leibler}, \citenamefont {Singh},\ and\ \citenamefont
	{Fisher}(1987)}]{Leibler:1987eq}%
\BibitemOpen
\bibfield  {author} {\bibinfo {author} {\bibfnamefont {S.}~\bibnamefont
		{Leibler}}, \bibinfo {author} {\bibfnamefont {R.}~\bibnamefont {Singh}}, \
	and\ \bibinfo {author} {\bibfnamefont {M.}~\bibnamefont {Fisher}},\ }\href
{\doibase 10.1103/PhysRevLett.59.1989} {\bibfield  {journal} {\bibinfo
		{journal} {Phys. Rev. Lett.}\ }\textbf {\bibinfo {volume} {59}},\ \bibinfo
	{pages} {1989} (\bibinfo {year} {1987})}\BibitemShut {NoStop}%
\bibitem [{\citenamefont {Halverson}\ \emph {et~al.}(2014)\citenamefont
	{Halverson}, \citenamefont {Smrek}, \citenamefont {Kremer},\ and\
	\citenamefont {Grosberg}}]{Halverson:2014hg}%
\BibitemOpen
\bibfield  {author} {\bibinfo {author} {\bibfnamefont {J.~D.}\ \bibnamefont
		{Halverson}}, \bibinfo {author} {\bibfnamefont {J.}~\bibnamefont {Smrek}},
	\bibinfo {author} {\bibfnamefont {K.}~\bibnamefont {Kremer}}, \ and\ \bibinfo
	{author} {\bibfnamefont {A.~Y.}\ \bibnamefont {Grosberg}},\ }\href {\doibase
	10.1088/0034-4885/77/2/022601} {\bibfield  {journal} {\bibinfo  {journal}
		{Rep. Prog. Phys.}\ }\textbf {\bibinfo {volume} {77}},\ \bibinfo {pages}
	{022601} (\bibinfo {year} {2014})}\BibitemShut {NoStop}%
\bibitem [{\citenamefont {Duplantier}(1987)}]{Duplantier:1987vw}%
\BibitemOpen
\bibfield  {author} {\bibinfo {author} {\bibfnamefont {B.}~\bibnamefont
		{Duplantier}},\ }\href
{http://eutils.ncbi.nlm.nih.gov/entrez/eutils/elink.fcgi?dbfrom=pubmed&id=9940716&retmode=ref&cmd=prlinks}
{\bibfield  {journal} {\bibinfo  {journal} {Phys. Rev. B}\ }\textbf {\bibinfo
		{volume} {35}},\ \bibinfo {pages} {5290} (\bibinfo {year}
	{1987})}\BibitemShut {NoStop}%
\end{thebibliography}

\end{document}